\documentclass[aps,preprint]{revtex4}%
\usepackage{amsfonts}
\usepackage{amsmath}
\usepackage{amssymb}
\usepackage[colorlinks,linkcolor=blue,citecolor=blue,urlcolor=blue]{hyperref}
\usepackage{subfigure}
\usepackage{graphicx}
\usepackage{cleveref}
\usepackage{listings}
\usepackage{xcolor}
\usepackage{array}%
\begin{document}
\preprint{CTP-SCU/2020022}
\title{Minimal Length Effects on Motion of a Particle in Rindler Space}
\author{Xiaobo Guo$^{a,b}$}
\email{guoxiaobo@czu.edu.cn}
\author{Kangkai Liang$^{d,e}$}
\email{lkk@berkeley.edu}
\author{Benrong Mu$^{c}$}
\email{benrongmu@cdutcm.edu.cn}
\author{Peng Wang$^{d}$}
\email{pengw@scu.edu.cn}
\author{Mingtao Yang$^{d}$}
\email{2017141221040@stu.scu.edu.cn}
\affiliation{$^{a}$Mechanical and Electrical Engineering School, Chizhou University,
Chizhou, Anhui, 247000, PR China}
\affiliation{$^{b}$Interdisciplinary Research Center of Quantum and Photoelectric
Information, Chizhou University, Chizhou, Anhui, 247000, PR China}
\affiliation{$^{c}$Physics Teaching and Research Section, College of Medical Technology,
Chengdu University of Traditional Chinese Medicine, Chengdu, 611137, PR China }
\affiliation{$^{d}$Center for Theoretical Physics, College of Physics, Sichuan University,
Chengdu, 610064, PR China}
\affiliation{$^{e}$Department of Physics, University of California at Berkeley, Berkeley,
CA, 94720, USA}

\begin{abstract}
Various quantum theories of gravity predict the existence of a minimal
measurable length. In this paper, we study effects of the minimal length on
the motion of a particle in the Rindler space under a harmonic potential. This
toy model captures key features of particle dynamics near a black hole
horizon, and allows us to make three observations. First, we find that the
chaotic behavior is stronger with the increases of the minimal length effects,
which manifests that the maximum Lyapunov characteristic exponents mostly
grow, and the KAM curves on Poincar\'{e} surfaces of section tend to
disintegrate into chaotic layers. Second, in the presence of the minimal
length effects, it can take a finite amount of Rindler time for a particle to
cross the Rindler horizon, which implies a shorter scrambling time of black
holes. Finally, it shows that some Lyapunov characteristic exponents can be
greater than the surface gravity of the horizon, violating the recently
conjectured universal upper bound. In short, our results reveal that quantum
gravity effects may make black holes prone to more chaos and faster scrambling.

\end{abstract}
\maketitle
\tableofcontents

\section{Introduction}

Nonlinear systems endowed with deterministic nature may behave in a
complicated, highly unpredictable and \textquotedblleft
chaotic\textquotedblright\ way. General relativity is a nonlinear dynamical
theory, and chaotic behavior in general relativity has been extensively
studied, e.g., chaoticity of cosmological solutions
\cite{Barrow:1981sx,Motter:2000bg}. Among various dynamical systems
investigated in general relativity, the test motion in a given black hole
spacetime is a quite hot topic in the literature, since it is of astrophysical
relevance and provides some important insights into AdS/CFT correspondence.

However, it is well known that the geodesic motion of a point particle in the
generic Kerr-Newman black hole spacetime is fully integrable
\cite{Carter:1968rr}. To induce chaos, one can resort to spacetimes with more
complicated geometries, external potentials imposed on test bodies,
perturbations introduced to backgrounds, or test bodies endowed with internal
structure. For a point particle, chaotic behavior of the geodesic motion has
been investigated in several static axisymmetric spacetimes \cite{Sota:1995ms}%
, multi-black hole spacetimes \cite{Hanan:2006uf}, bumpy spacetimes
\cite{Gair:2007kr}, weakly magnetized Schwarzschild black holes
\cite{Zahrani:2013up}, black holes with discs or rings \cite{Witzany:2015yqa},
the Schwarzschild--Melvin black holes \cite{Wang:2016wcj}, accelerating black
holes \cite{Chen:2016tmr} and spacetimes with a quadrupole mass moment
\cite{Wang:2018eui}. In a universal way, the particle chaotic motion has
lately been studied near the black hole horizon
\cite{Hashimoto:2016dfz,Dalui:2019umw,Dalui:2019esx}. Interestingly,
gravitational waves emitted from chaotic motions of particles in the
bumpy\ spacetime can be used to distinguish an extreme-mass-ratio inspiral
into a Kerr background spacetime from one into a non-Kerr background spacetime
\cite{Apostolatos:2009vu}. Recently, it has been shown that such proposal may
be undermined due to chaos suppression by frame dragging
\cite{Gutierrez-Ruiz:2018tre}. Partly motivated by AdS/CFT correspondence,
chaotic dynamics of a ring string has been studied in various black hole
backgrounds
\cite{Zayas:2010fs,Ma:2014aha,Basu:2016zkr,Hashimoto:2018fkb,Cubrovic:2019qee,Ma:2019ewq}
since the geodesic motion of a ring string was shown to exhibit chaotic
behavior in a Schwarzschild black hole \cite{Frolov:1999pj}. In addition, as a
simplified model describing extreme mass ratio inspirals, the motion of a
spinning test particle in black hole backgrounds was considered and also
demonstrated to possess some chaotic features
\cite{Suzuki:1996gm,Hartl:2003da,Lukes-Gerakopoulos:2016bup,Zelenka:2019nyp}.

On the other hand, the existence of a minimal measurable length has been
predicted in various quantum theories of gravity such as string theory
\cite{Veneziano:1986zf,Gross:1987ar,Amati:1988tn,Garay:1994en,Scardigli:1999jh}%
. To incorporate the minimal length into quantum mechanics, the Heisenberg
uncertainty principle can be modified, giving the so called \textquotedblleft
the generalized uncertainty principle (GUP)\textquotedblright%
\ \cite{Maggiore:1993kv,Kempf:1994su}. Usually, the fundamental commutation
relation is deformed to realize the GUP. For a $1$D quantum system, the
deformed commutator between position and momentum can take the following form
\begin{equation}
\lbrack X,P]=i\hbar(1+\beta P^{2}), \label{eq:1dGUP}%
\end{equation}
where $\beta$ is some deformation parameter, and the minimal length is $\Delta
X_{\min}=\hbar\sqrt{\beta}$. Many minimal length deformed quantum systems have
been investigated intensively in the literature, e.g., the harmonic oscillator
\cite{Chang:2001kn}, Coulomb potential \cite{Akhoury:2003kc,Brau:1999uv},
gravitational well \cite{Brau:2006ca,Pedram:2011xj}, quantum optics
\cite{Pikovski:2011zk,Bosso:2018ckz}, compact stars
\cite{Wang:2010ct,Wang:2011iv,Ong:2018zqn} and cosmology
\cite{Guo:2016btf,Khodadi:2018wed,Khodadi:2018scn}. Furthermore, taking the
classical limit $\hbar\rightarrow0$, one can discuss the minimal length
effects on classical systems, such as observational tests of general
relativity
\cite{Benczik:2002tt,Ahmadi:2014cga,Silagadze:2009vu,Scardigli:2014qka,Ali:2015zua,Guo:2015ldd,Khodadi:2017eim,Scardigli:2018qce}%
, classical harmonic oscillator \cite{Tao:2012fp,Quintela:2015bua},
equivalence principle \cite{Tkachuk:2013qa}, Newtonian potential
\cite{Scardigli:2016pjs}, the Schr\"{o}dinger-Newton equation
\cite{Zhao:2017xjj}, the weak cosmic censorship conjecture \cite{Mu:2019bim}
and motions of particles near a black hole horizon
\cite{Lu:2018mpr,Hassanabadi:2019iff,Maghsoodi:2020ura}. In addition, the
minimal length corrected Hawking temperature can be obtained by using the
Hamilton-Jacobi method
\cite{Chen:2013pra,Chen:2013tha,Chen:2013ssa,Chen:2014xgj,Maghsoodi:2019fca}.

In this paper, we discuss the minimal length effects on the motion of a
particle under a harmonic potential in the Rindler space, which is the
approximation of the near-horizon region. This study is a follow-up of our
previous works \cite{Lu:2018mpr,Guo:2020xnf}, which demonstrated that the
minimal length effects tend to increase chaos. Analytical approaches, i.e.,
the perturbation method and Melnikov method, were employed to investigate the
chaotic motion of a particle around a black hole in
\cite{Lu:2018mpr,Guo:2020xnf}. Here, we numerically study the motion of a
particle and the corresponding chaos indicators in the Rindler space. Our
numerical results not only support the findings of
\cite{Lu:2018mpr,Guo:2020xnf}, but also signal a shorter scrambling time, a
notion that is connected to chaos.

The rest of this paper is organized as follows. In section \ref{Sec:MPRS}, we
obtain the equations of motion for the dynamical system. The dynamics of the
system is numerically analyzed in section \ref{Sec:DPM}. We summarize our
results with a brief discussion in section \ref{Sec:DC}. In this paper, we
take Geometrized units $c=G=k_{B}=1$, where the Planck constant $\hbar$ is
square of the Planck length $\ell_{p}$.

\section{Motion of a Particle in Rindler Space}

\label{Sec:MPRS}

To study the minimal length effects on chaotic dynamics of a particle, we
consider the motion of the particle in the near-horizon region, where chaotic
behavior can be induced. Specifically, we discuss a relativistic particle
moving in the near-horizon region of a $4$D spherically symmetric black hole
with the metric%
\begin{equation}
ds^{2}=-h\left(  r\right)  dt^{2}+\frac{dr^{2}}{g\left(  r\right)  }%
+r^{2}\left(  d\theta^{2}+\sin^{2}\theta d\phi^{2}\right)  , \label{eq:BH}%
\end{equation}
where $h\left(  r\right)  $ and $g\left(  r\right)  $ are assumed to have a
simple zero at the event horizon $r=r_{+}$. The Hawking temperature is
\begin{equation}
T=\frac{\hbar\sqrt{g^{\prime}\left(  r_{+}\right)  h^{\prime}\left(
r_{+}\right)  }}{4\pi}\equiv\frac{\hbar\alpha}{2\pi}, \label{eq:ht}%
\end{equation}
where we define the surface gravity $\alpha\equiv\sqrt{g^{\prime}\left(
r_{+}\right)  h^{\prime}\left(  r_{+}\right)  }/2$ for later use. To explore
the region near the horizon, we introduce the proper distance from the
horizon,%
\begin{equation}
x=\int_{r_{+}}^{r}\frac{dr}{\sqrt{g\left(  r\right)  }}\sim\frac
{2\sqrt{r-r_{+}}}{\sqrt{g^{\prime}\left(  r_{+}\right)  }}.
\end{equation}
If one focuses on a small angular near-horizon region centered at $\theta=0$,
the near-horizon metric of the black hole can be rewritten in terms of $x$,%
\begin{equation}
ds^{2}=-\alpha^{2}x^{2}dt^{2}+dx^{2}+dy^{2}+dz^{2},
\label{eq:RindlerCoordinates}%
\end{equation}
where we introduce Cartesian coordinates,%
\begin{equation}
y=r_{+}\theta\cos\phi\text{ and }z=r_{+}\theta\sin\phi.
\end{equation}
The metric $\left(  \ref{eq:RindlerCoordinates}\right)  $ is the Rindler
space, which describes the near-horizon geometry of the black hole $\left(
\ref{eq:BH}\right)  $. Note that the black hole horizon is at $x=0$ in the
Rindler coordinates. Since we are interested in the motion of a particle in
the near-horizon region, we confine ourselves here to considering the Rindler space.

It is well known that the geodesic equation in the metric $\left(
\ref{eq:BH}\right)  $ is separable, and hence the geodesic motion of a
particle is integrable. To make the motion of a particle chaotic, one can
impose an external potential outside the horizon. In what follows, we assume
that the potential is a harmonic potential centered at $(x,y,z)=(x_{0},0,0)$
with $x_{0}>0$,
\begin{equation}
V\left(  x,y,z\right)  =\frac{\omega^{2}}{2}\left[  \left(  x-x_{0}\right)
^{2}+y^{2}+z^{2}\right]  ,
\end{equation}
where $\omega$ is the angular frequency.

Incorporating the deformed fundamental commutation relation $\left(
\ref{eq:1dGUP}\right)  $, the minimal length deformed Hamilton-Jacobi equation
for the motion of a particle under an external potential has been derived in
\cite{Lu:2018mpr}. In the Rindler space with the harmonic potential $V\left(
x,y,z\right)  $, the deformed Hamilton-Jacobi equation then becomes%
\begin{equation}
\frac{1}{\alpha^{2}x^{2}}\left[  \frac{\partial S}{\partial t}+V\left(
x,y,z\right)  \right]  ^{2}-\mathcal{X}\left(  1+\beta\mathcal{X}\right)
^{2}=m^{2}, \label{eq:HJE}%
\end{equation}
where $\mathcal{X}\equiv\left(  \partial_{x}S\right)  ^{2}+\left(
\partial_{y}S\right)  ^{2}+\left(  \partial_{z}S\right)  ^{2}$, and $S$ is the
classical action. The first derivatives of $S$ with respect to the spatial
coordinates are the conjugate momenta,
\begin{equation}
p_{i}=\frac{\partial S}{\partial x_{i}}\text{ for }i=x,y,z,
\end{equation}
and the Hamiltonian of the system corresponds to the first derivative of $S$
with respect to the time,%
\begin{equation}
\mathcal{H}=-\frac{\partial S}{\partial t}.
\end{equation}
Solving the Hamilton-Jacobi equation $\left(  \ref{eq:HJE}\right)  $ for
$\mathcal{H}$ gives the Hamiltonian
\[
\mathcal{H}=\alpha x\sqrt{p^{2}\left(  1+\beta p^{2}\right)  ^{2}+m^{2}%
}+V\left(  x,y,z\right)  ,
\]
where we define
\begin{equation}
p^{2}\equiv p_{x}^{2}+p_{y}^{2}+p_{z}^{2}.
\end{equation}
The equations of motion for $x_{i}$ are%
\begin{equation}
\dot{x}_{i}=\frac{\partial\mathcal{H}}{\partial p_{i}}=\alpha xp_{i}%
\frac{\left(  1+\beta p^{2}\right)  \left(  1+3\beta p^{2}\right)  }%
{\sqrt{p^{2}\left(  1+\beta p^{2}\right)  ^{2}+m^{2}}}\text{ for }i=x,y,z,
\label{eq:eomx}%
\end{equation}
and those for $p_{i}$ are%
\begin{align}
\dot{p}_{x}  &  =-\frac{\partial\mathcal{H}}{\partial x}=-\alpha\sqrt
{p^{2}\left(  1+\beta p^{2}\right)  ^{2}+m^{2}}-\omega^{2}\left(
x-x_{0}\right)  ,\nonumber\\
\dot{p}_{y}  &  =-\frac{\partial\mathcal{H}}{\partial y}=-\omega^{2}y\text{,
}\dot{p}_{z}=-\frac{\partial\mathcal{H}}{\partial z}=-\omega^{2}z\text{.}
\label{eq:eomp}%
\end{align}
The equations of motion for $z$ and $p_{z}$ show that, a particle which starts
in the $z=0$ plane, remains in the $z=0$ plane. For simplicity, we only
consider the motion of a particle on the $z=0$ plane in this paper. Therefore,
we set $z=0=p_{z}$ in the remainder of the paper.

\section{Dynamics of Particle Motion}

\label{Sec:DPM}

\begin{figure}[tb]
\begin{center}
\includegraphics[width=0.5\textwidth]{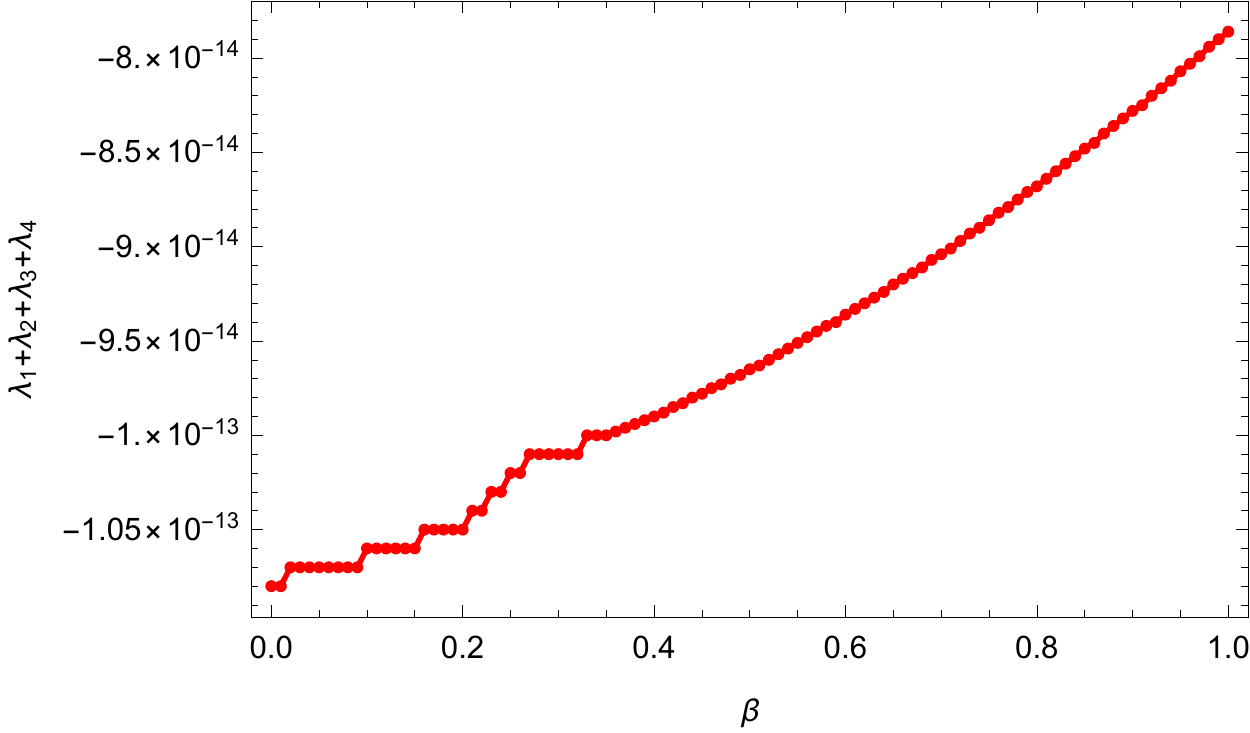}
\includegraphics[width=0.48\textwidth]{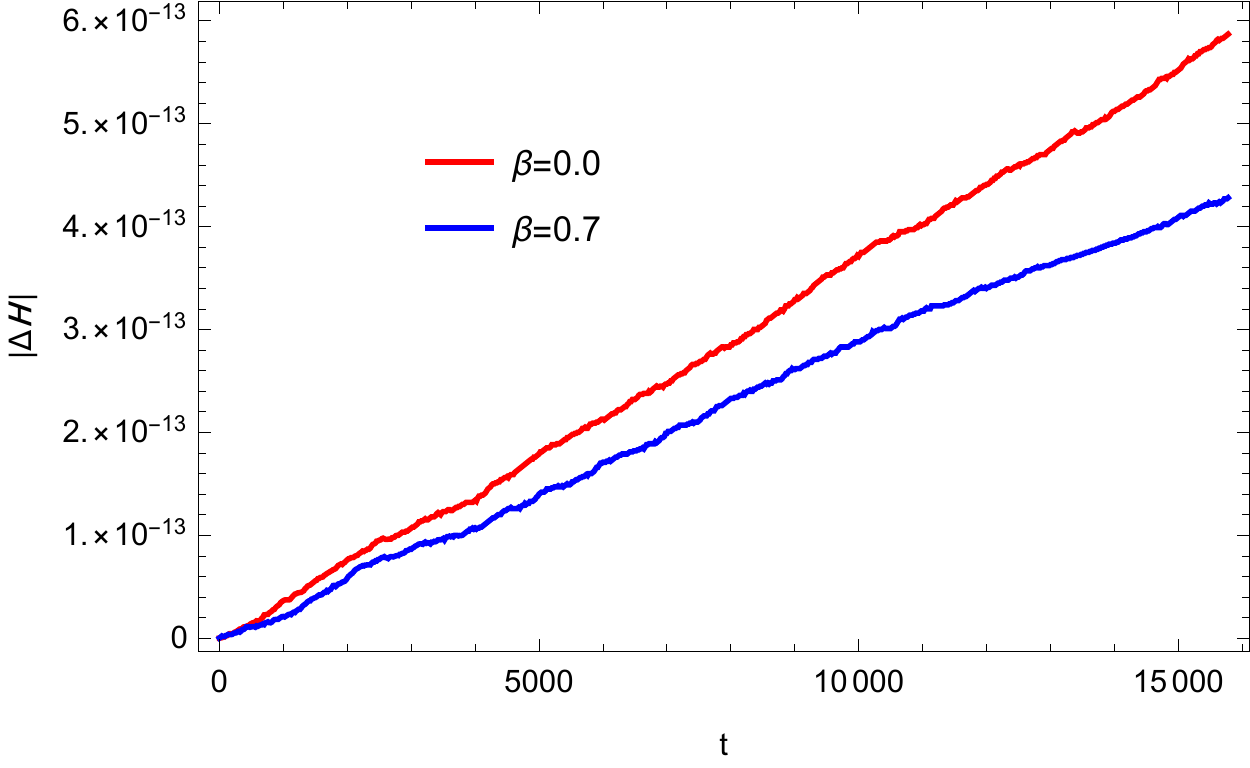}
\end{center}
\caption{{\footnotesize Numerical accuracy. \textbf{Left Panel}: Sum of LCEs
as a function of $\beta$ for $E=5.0$. \textbf{Right Panel}: Errors of energy
with time for $\beta=0$ and $0.7$. }}%
\label{fig:CheckA}%
\end{figure}

In this section, we study the dynamics of a particle moving in the Rindler
space under the external potential $V\left(  x,y,z\right)  $, especially the
minimal length effects on the chaotic dynamics. To detect chaotic phenomenon
in the dynamical system, we resort to Poincar\'{e} surfaces of section and
Lyapunov characteristic exponents (LCEs), which are calculated numerically.
Since the chaotic motion of a particle is very sensitive to initial values, a
numerical method with high precision is highly desirable. Here, we adopt to
Verner's \textquotedblleft most efficient\textquotedblright\ Runge-Kutta
$9(8)$ method \cite{VERNER1996345}, which can achieve high accuracy solving
(tolerances like $<10^{-12}$). To test the accuracy of the numerical method,
we consider two conserved quantities, namely the energy of the system $E$,
which is conserved since there is no dissipation, and the sum of all LCEs
$\sum\nolimits_{i=1}^{4}\lambda_{i}$, which must be zero since a volume
element of the phase space will stay the same along a trajectory for the
conservative system. The left panel of FIG. \ref{fig:CheckA} shows the sum of
all LCEs as a function of $\beta$ with the energy $E=5.0$. We also display the
error of the energy $\Delta\mathcal{H\equiv H}\left(  t\right)  $
$-\mathcal{H}\left(  0\right)  $ along two trajectories with $\beta=0$ and
$0.7$, respectively, in the right panel of FIG. \ref{fig:CheckA}. It is
exhibited that the numerical method used in this paper can maintain the
numerical error around or below $10^{-13}$. We choose $\alpha=1$, $\omega=10$,
$m=1$ and $x_{0}=1$ for the numerical analysis in this section.

\subsection{Fixed Point and Orbits}

Invariant set, such as fixed points and limit cycles, is an important concept
to understand late time dynamics and stability of a dynamical system. In
particular, we here find the fixed point in the phase space of the dynamical
system described by eqns. $\left(  \ref{eq:eomx}\right)  $ and $\left(
\ref{eq:eomp}\right)  $, and discuss the behavior near the corresponding fixed
point. The fixed point\ is an equilibrium point, which corresponds to $\dot
{x}_{i}=0=\dot{p}_{i}$. This leads to the fixed point solution%
\begin{equation}
x_{f}=x_{0}-\frac{m\alpha}{\omega^{2}}\text{, }y_{f}=0\text{, }p_{xf}=0\text{
and }p_{yf}=0. \label{eq:fp}%
\end{equation}
Since $x_{f}\geq0$ in the Rindler space, the fixed point disappears when
$x_{0}<$ $m\alpha/\omega^{2}$. Near the fixed point, the nonlinear system can
be approximated by a linear system represented by the Jacobian matrix, the
eigenvalues of which determine the near-fixed point behavior. The eigenvalues
of the Jacobian matrix at the fixed point $\left(  \ref{eq:fp}\right)  $ are
given by%
\begin{equation}
\left\{  i\omega\sqrt{\frac{x_{f}\alpha}{m}},i\omega\sqrt{\frac{x_{f}\alpha
}{m}},-i\omega\sqrt{\frac{x_{f}\alpha}{m}},-i\omega\sqrt{\frac{x_{f}\alpha}%
{m}}\right\}  .
\end{equation}
Since the eigenvalues are purely imaginary, this fixed point is a center in
the sense that trajectories near the fixed point are almost closed loops.

\begin{figure}[tb]
\begin{center}
\includegraphics[width=0.242\textwidth]{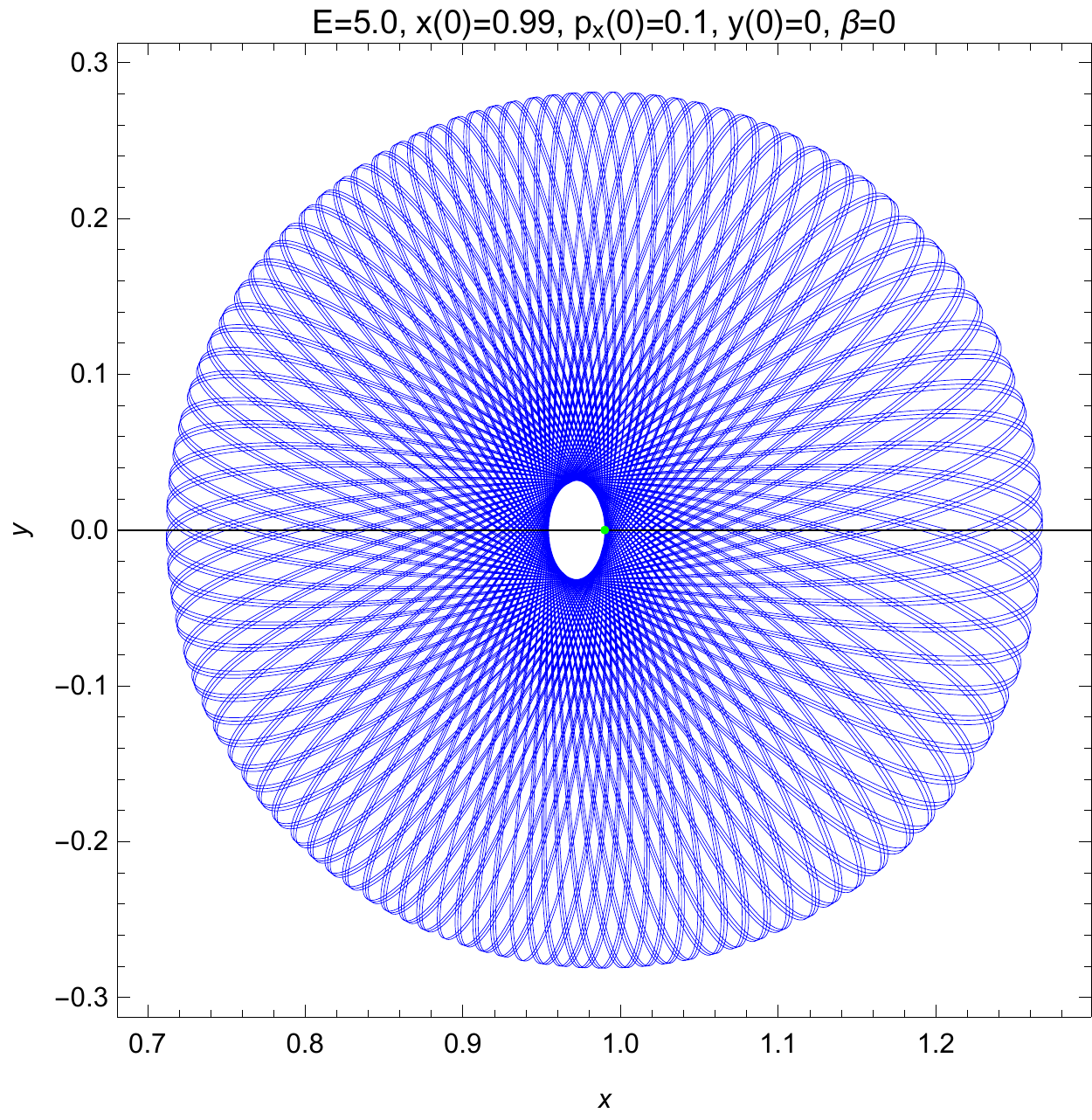}
\includegraphics[width=0.242\textwidth]{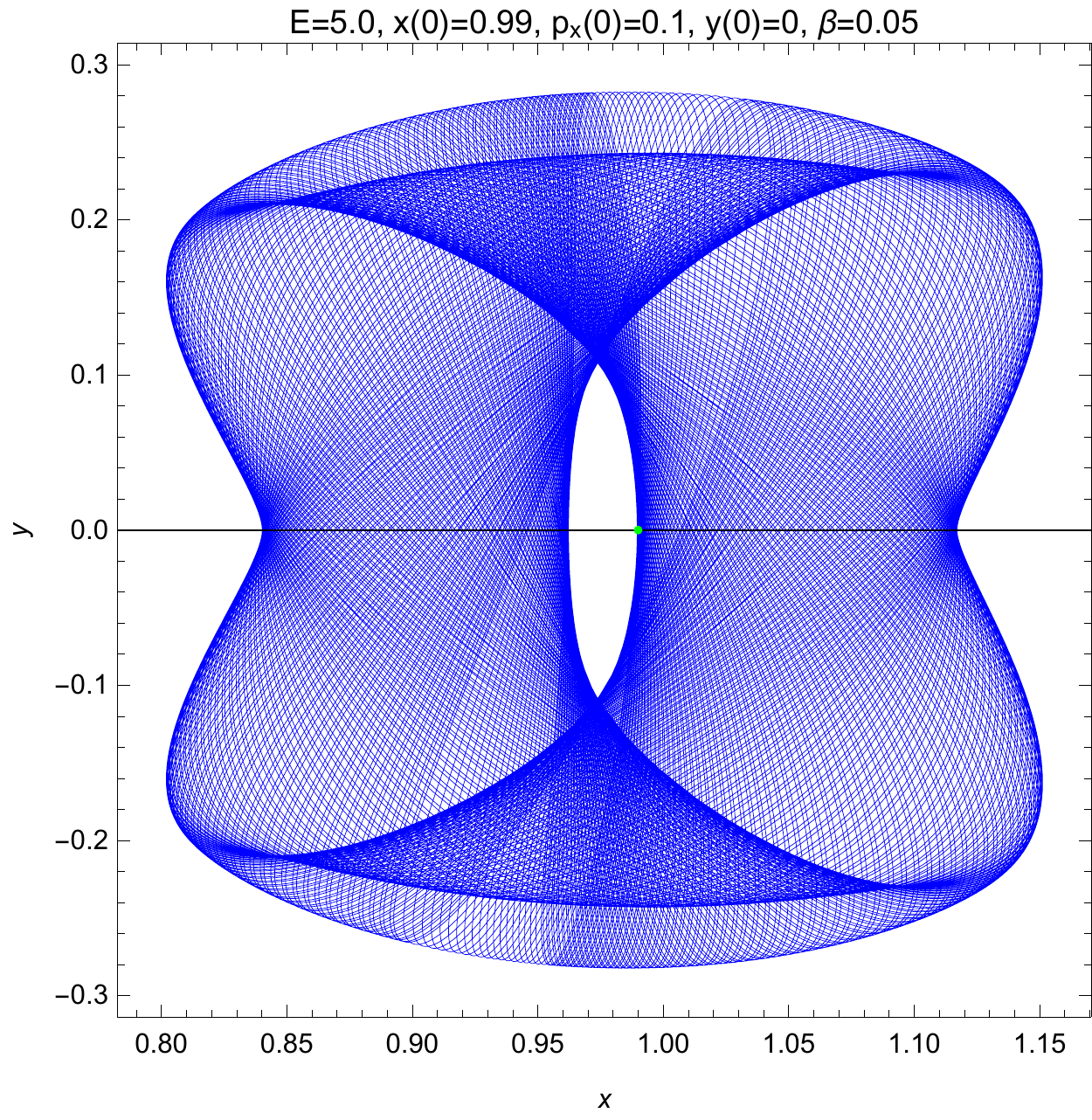}
\includegraphics[width=0.242\textwidth]{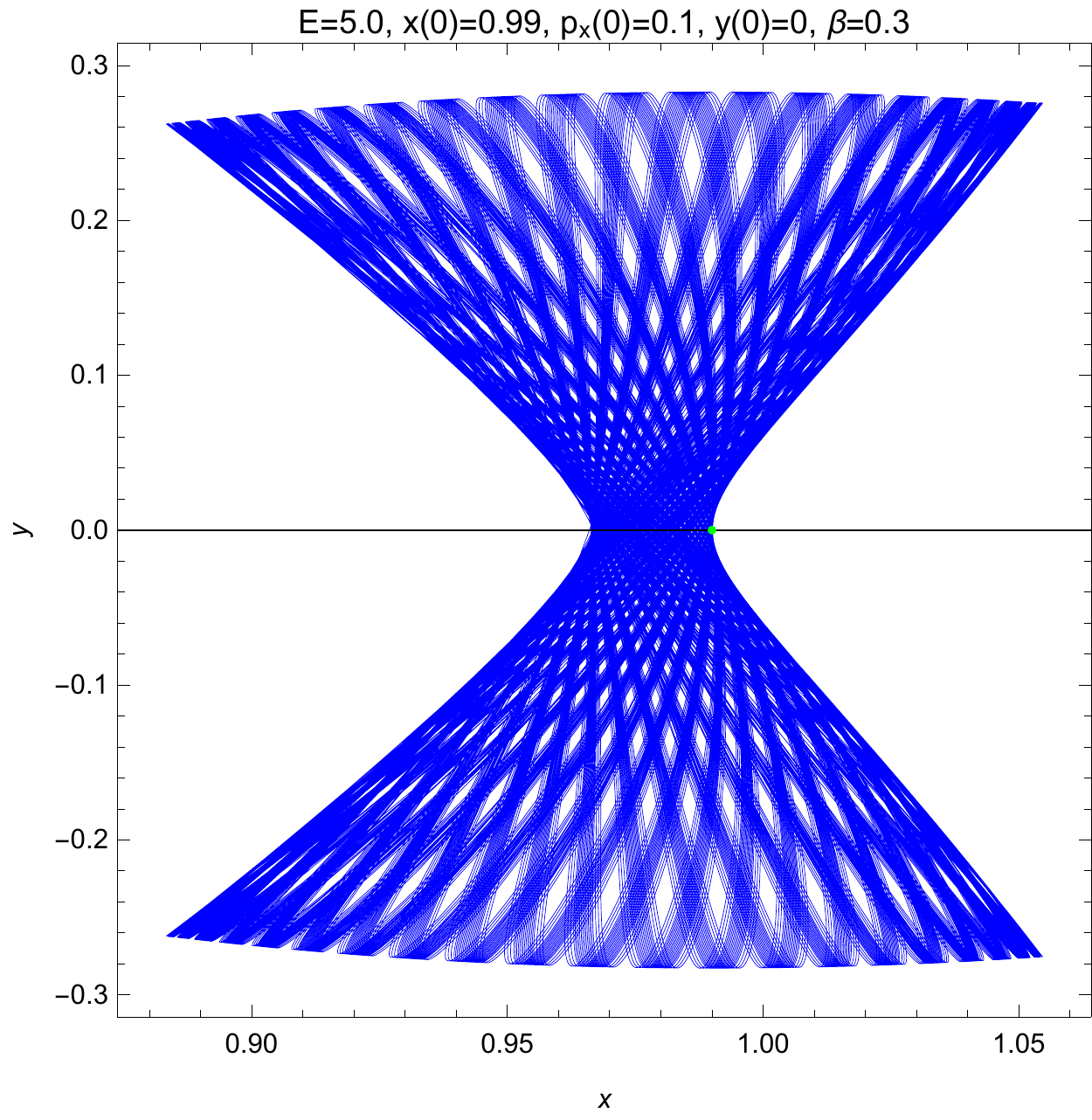}
\includegraphics[width=0.242\textwidth]{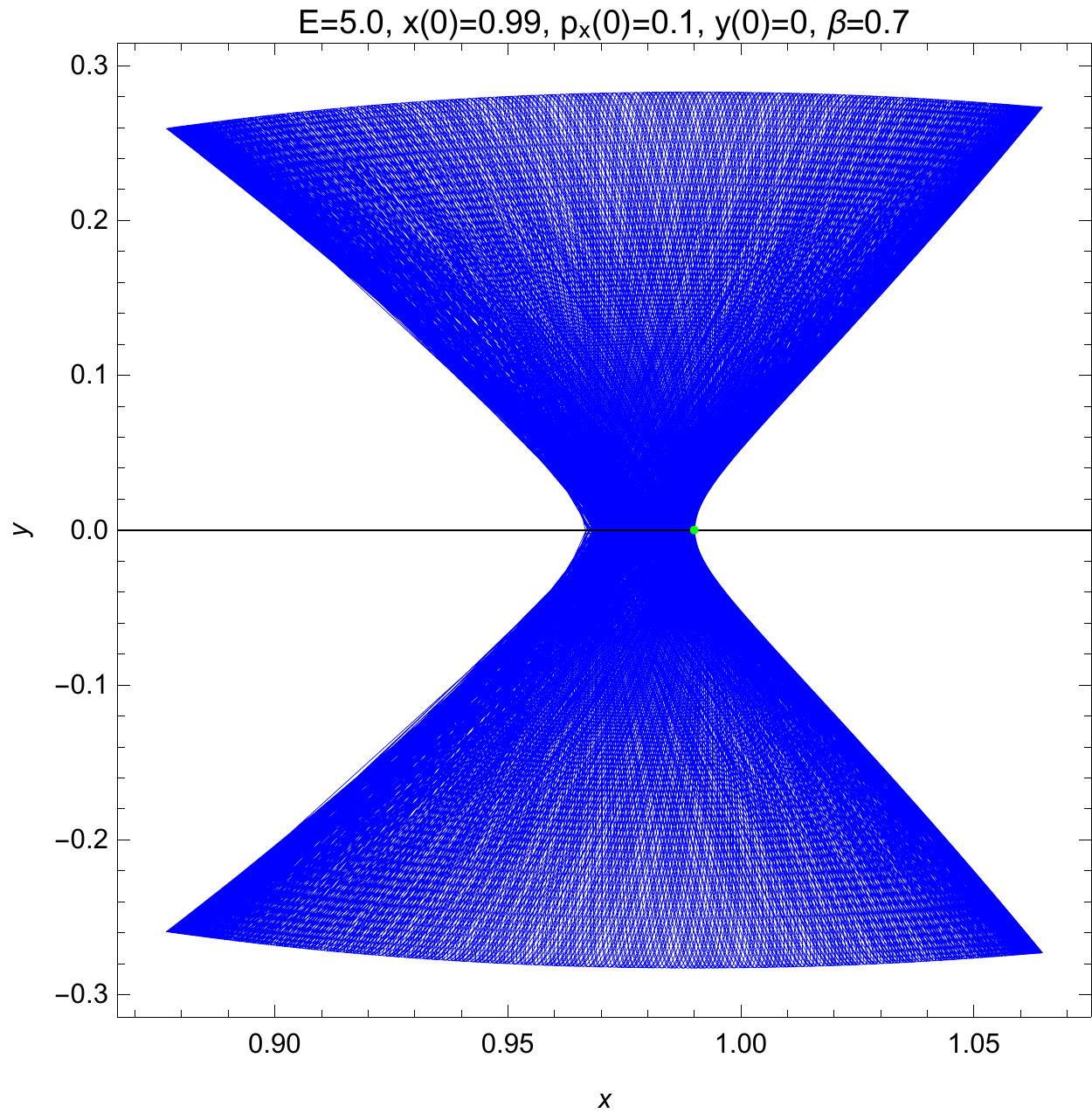}
\end{center}
\caption{{\footnotesize Orbits in the $x$-$y$ plane with the initial
conditions $E=5.0$, $x\left(  0\right)  =0.99$, $p_{x}\left(  0\right)  =0.1$
and $y\left(  0\right)  =0$ for $\beta=0$, $0.05$, $0.3$ and $0.7$. The orbit
tends to be more irregular for a larger value of $\beta$. The green dots
represent the starting points of the orbits, which are the fixed point.}}%
\label{fig:TraE5}%
\end{figure}

\begin{figure}[tb]
\begin{center}
\includegraphics[width=0.242\textwidth]{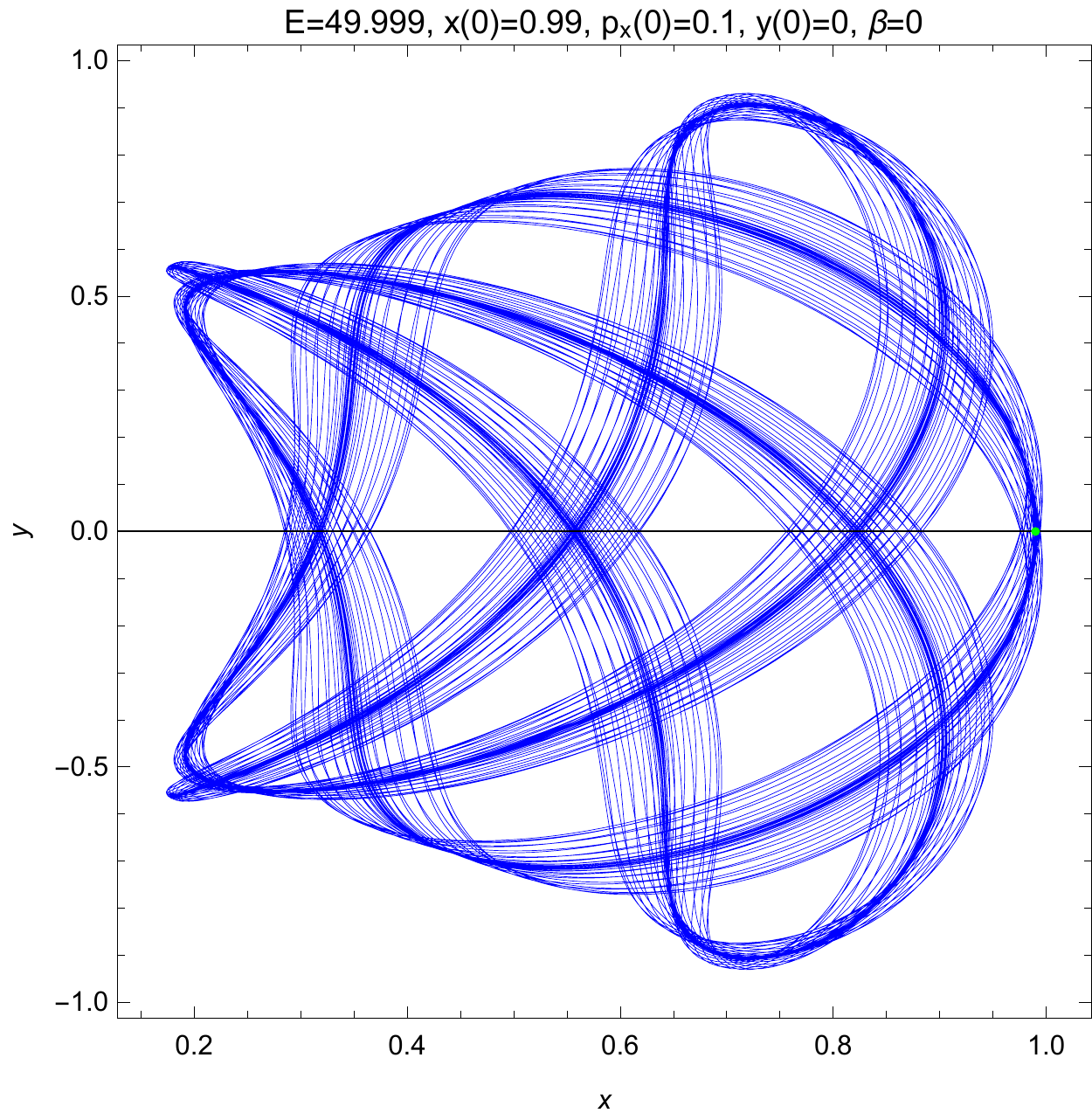}
\includegraphics[width=0.242\textwidth]{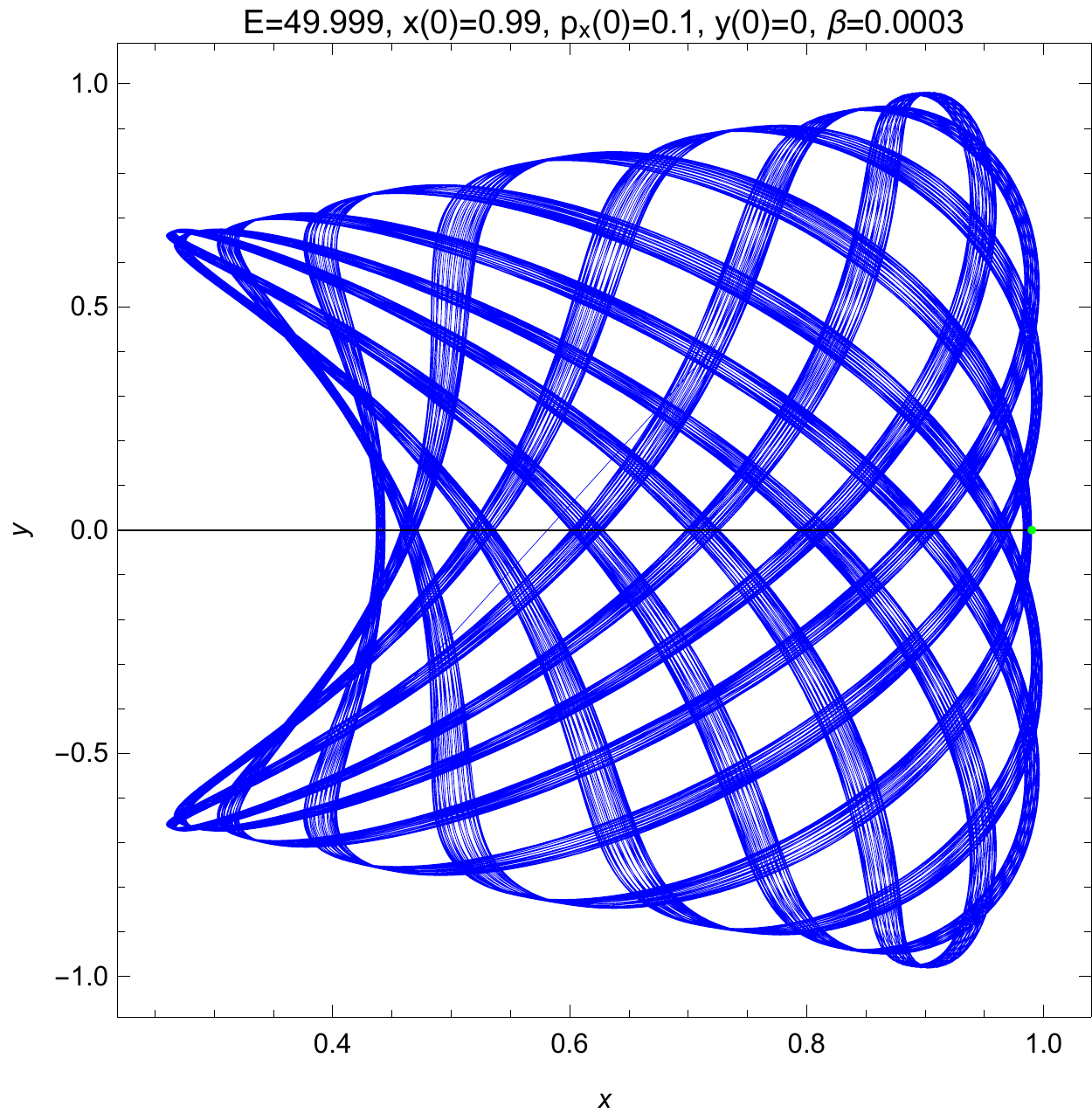}
\includegraphics[width=0.242\textwidth]{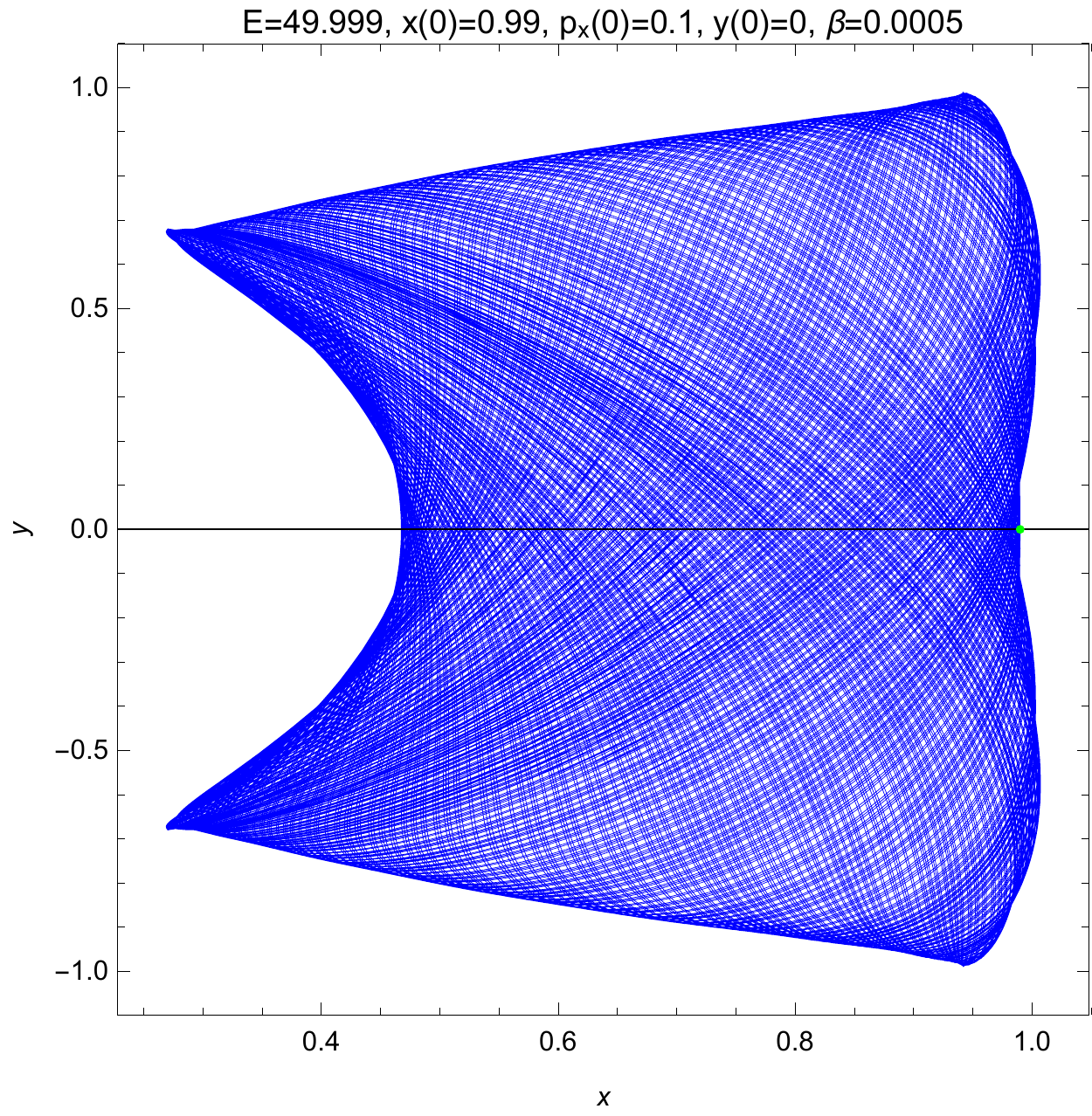}
\includegraphics[width=0.242\textwidth]{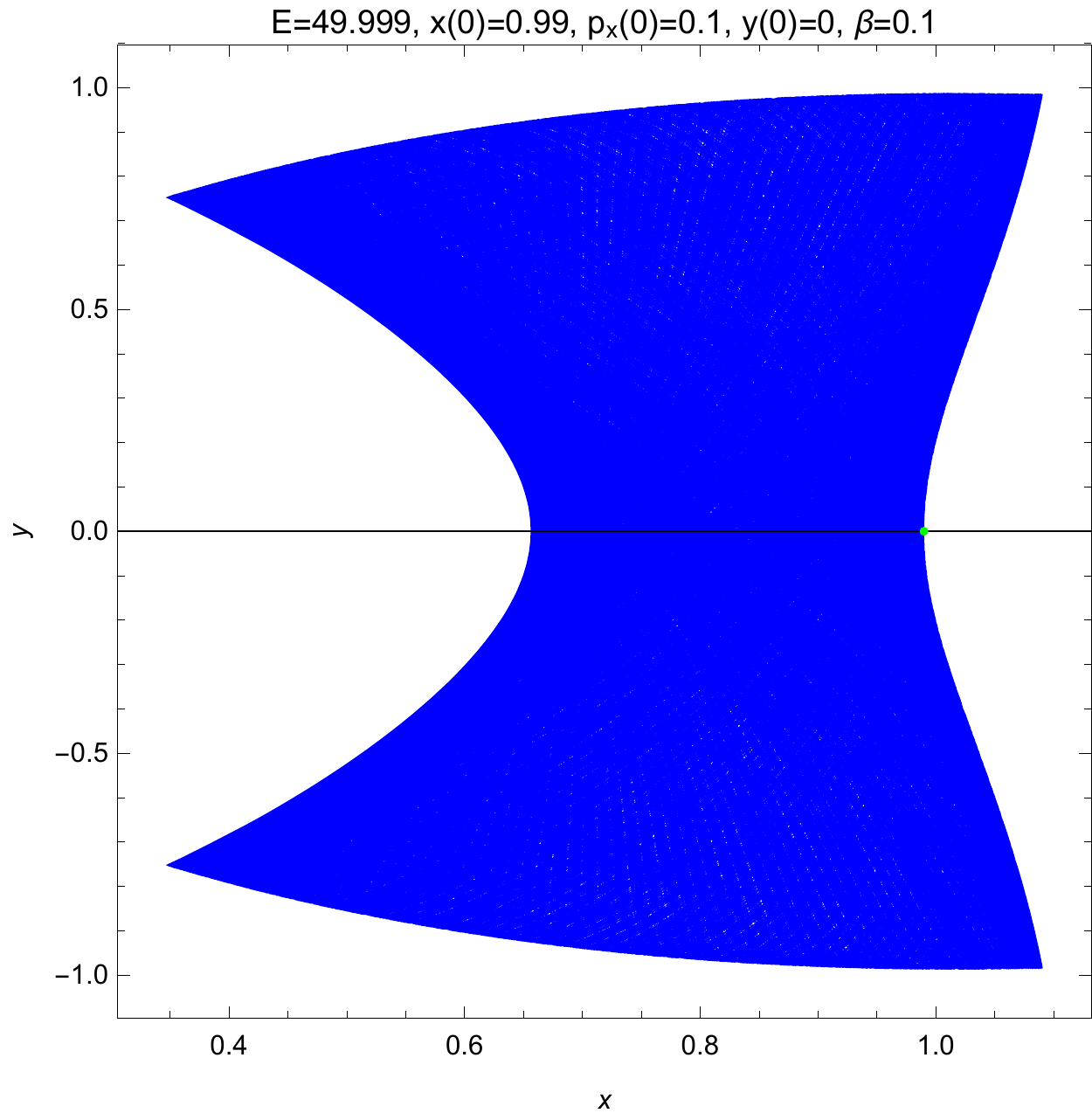}
\end{center}
\caption{{\footnotesize Orbits in the $x$-$y$ plane with the initial
conditions $E=49.999$, $x\left(  0\right)  =0.99$, $p_{x}\left(  0\right)
=0.1$ and $y\left(  0\right)  =0$ for $\beta=0$, $0.0003$, $0.0005$ and $0.1$.
As $\beta$ increases, the orbit is likely to be more erratic. The green dots
represent the starting points of the orbits, which are the fixed point.}}%
\label{fig:TraE49}%
\end{figure}

The fixed point solution $\left(  \ref{eq:fp}\right)  $ corresponds to the
minimum energy of the system, $E_{\min}=m\alpha x_{0}\left(  1-\frac
{ma}{2\omega^{2}x_{0}}\right)  $. At the event horizon located at $x=0$, the
energy of the system $E$ should satisfy%
\[
E\geq E_{\max}\equiv V\left(  0,0,0\right)  =\frac{\omega^{2}x_{0}^{2}}{2}.
\]
Hence for $E_{\min}<E<E_{\max}$, the system wanders around the fixed point,
and can never reach or cross the event horizon. In FIG. \ref{fig:TraE5}, we
present the orbits of the system in the $x$-$y$ plane with the initial values
$E=5.0$ ($>E_{\min}=0.995$), $x\left(  0\right)  =0.99$, $p_{x}\left(
0\right)  =0.1$ and $y\left(  0\right)  =0$ for $\beta=0$, $0.05$, $0.3$ and
$0.7$. When $\beta=0$, the orbit appears to be regular and periodic. With the
increasing value of $\beta$, the orbit starts to become irregular.
Particularly, the orbit in the $\beta=0.7$ case is shown to be quite erratic.
FIG. \ref{fig:TraE49} displays the orbits in the $x$-$y$ plane with the
initial values $E=49.999$ ($<E_{\max}=50.0$), $x\left(  0\right)  =0.99$,
$p_{x}\left(  0\right)  =0.1$ and $y\left(  0\right)  =0$ for $\beta=0$,
$0.0003$, $0.0005$ and $0.1$. It is noteworthy that the system with $\beta=0$
already exhibits irregular movement. When $\beta$ increases, the orbit becomes
more irregular. These observations signal that the dynamical system becomes
more chaotic as $\beta$ increases. \begin{figure}[tb]
\begin{center}
\includegraphics[width=0.28\textwidth]{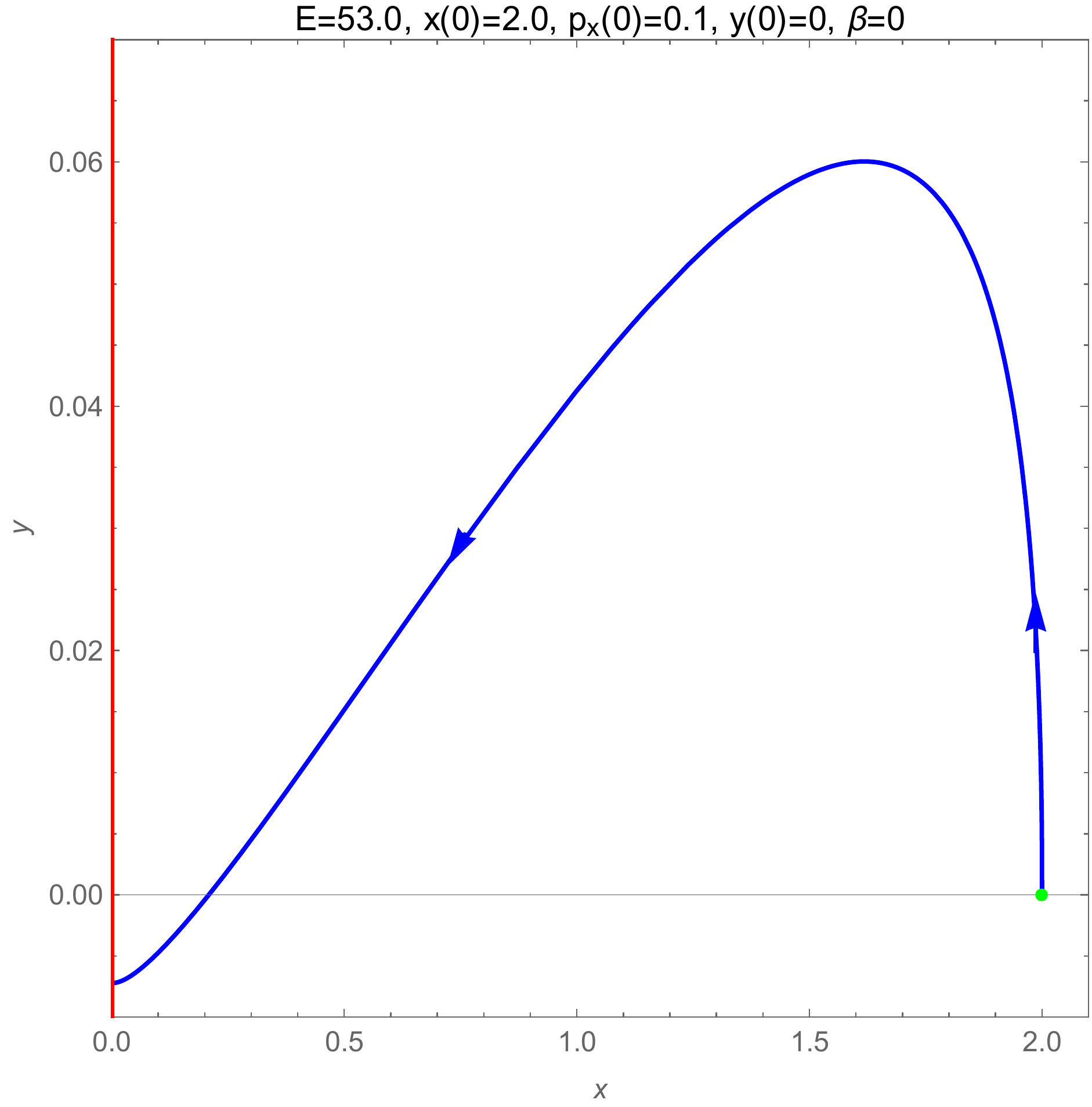}
\includegraphics[width=0.35\textwidth]{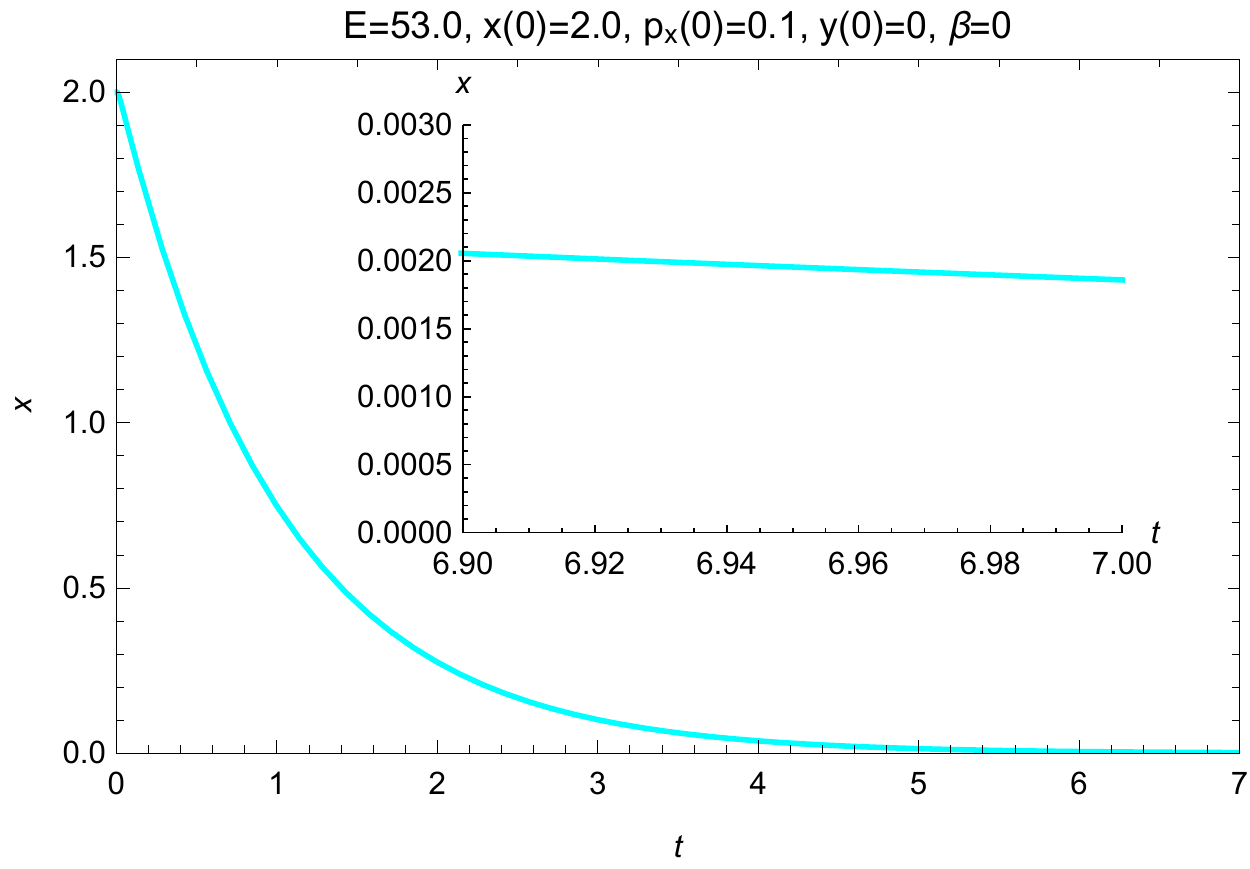}
\includegraphics[width=0.35\textwidth]{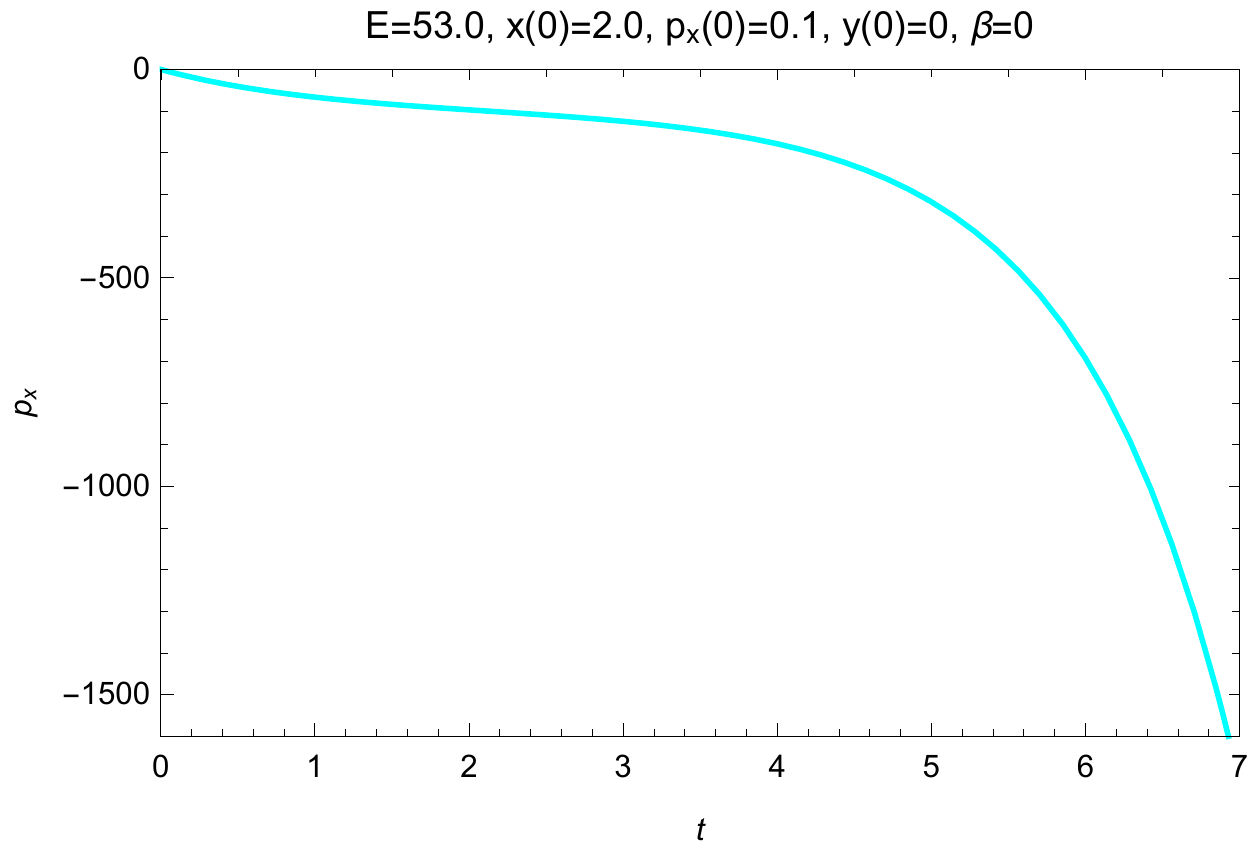}
\end{center}
\caption{{\footnotesize Plots showing the orbit in the $x$-$y$ plane,
$x\left(  t\right)  $ and $p_{x}\left(  t\right)  $ for $\beta=0$ with the
initial conditions $E=53.0$, $x\left(  0\right)  =2.0$, $p_{x}\left(
0\right)  =0.1$ and $y\left(  0\right)  =0$. The red line and green dot
represent the event horizon and the staring point of the orbit, respectively.
The arrows show the increase in simulation time. The proper spatial of the
particle from the horizon exponentially decreases with time, whereas the
magnitude of the momentum $p_{x}$ grows exponentially with time.}}%
\label{fig:TraE53beta0}%
\end{figure}

\begin{figure}[tb]
\begin{center}
\includegraphics[width=0.28\textwidth]{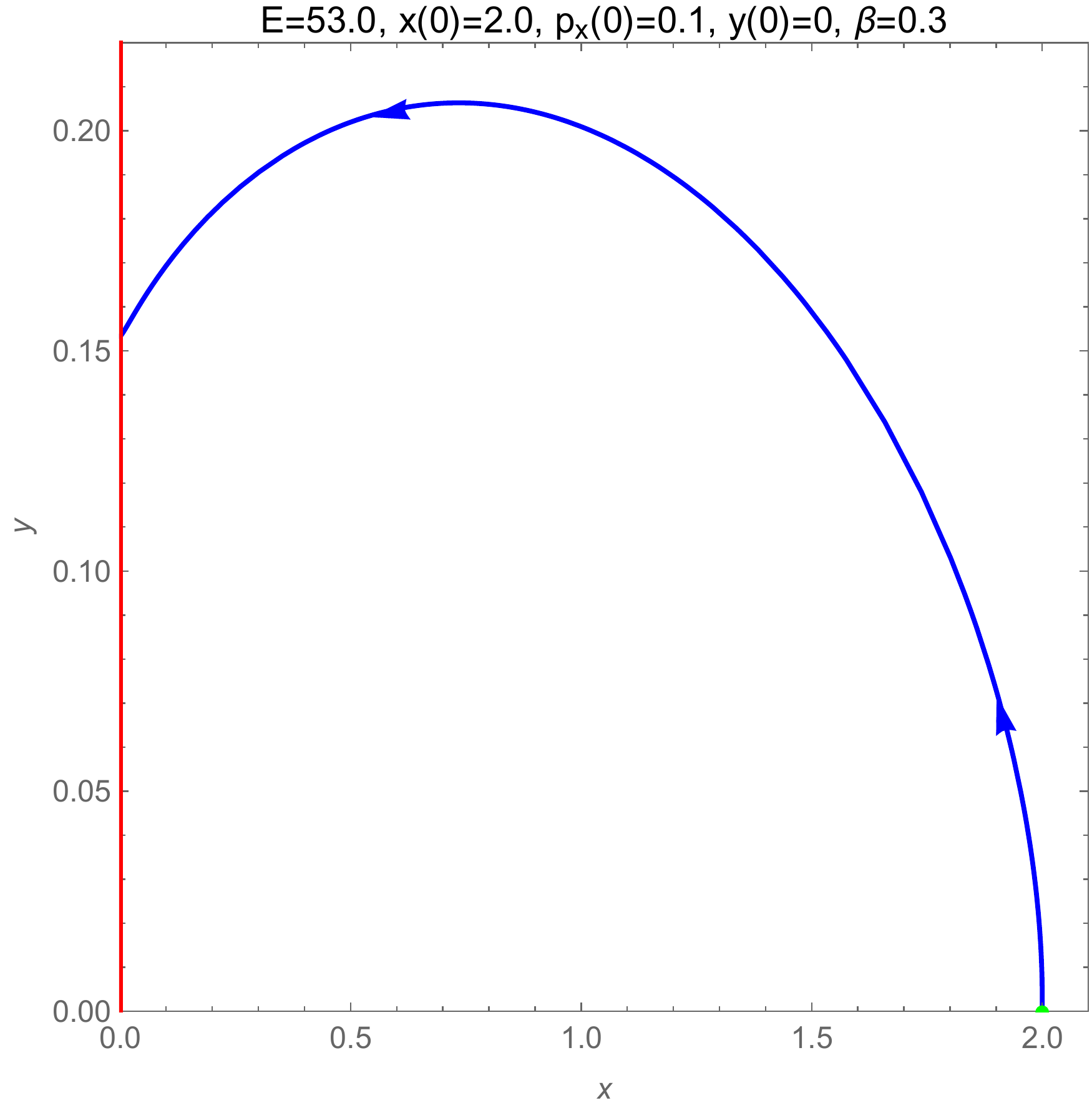}
\includegraphics[width=0.35\textwidth]{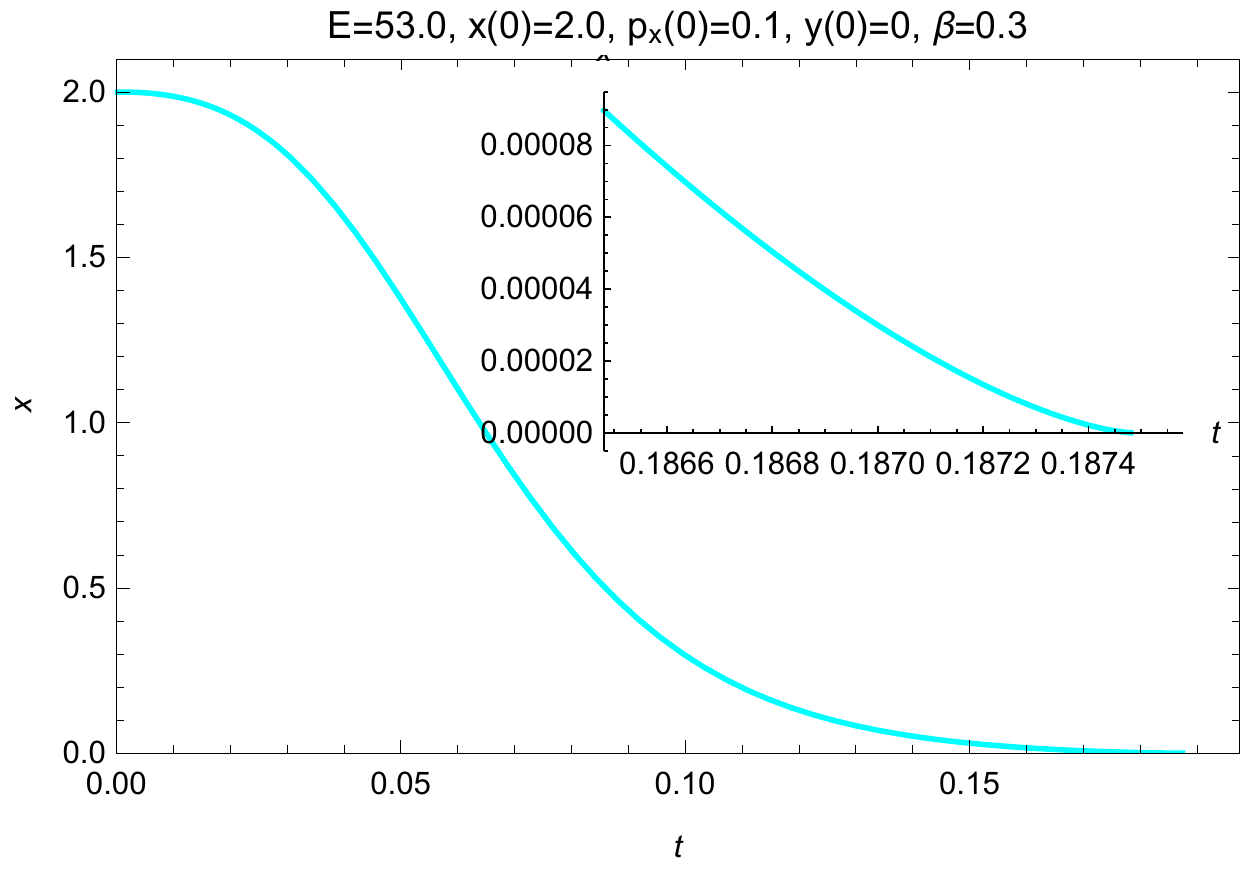}
\includegraphics[width=0.35\textwidth]{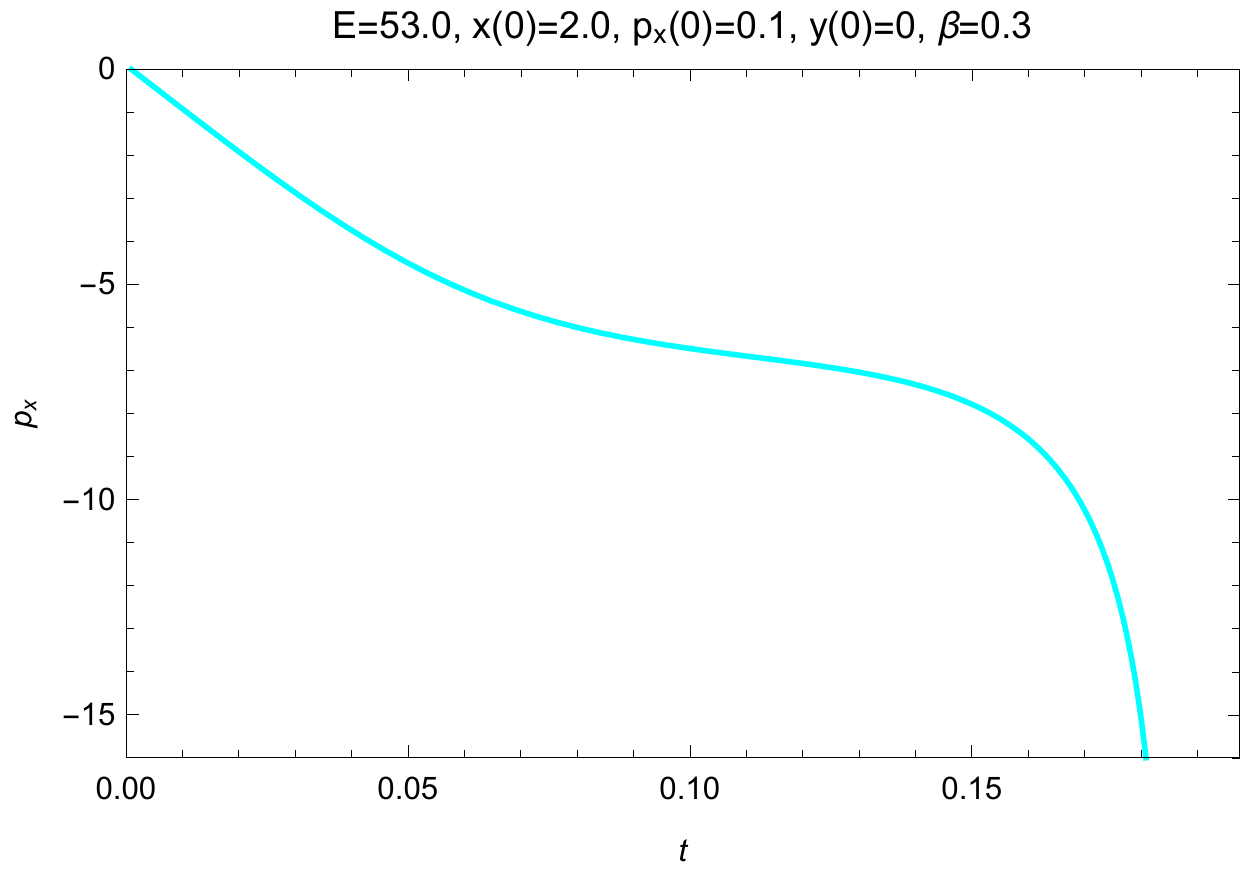}
\end{center}
\caption{{\footnotesize Plots showing the orbit in the $x$-$y$ plane,
$x\left(  t\right)  $ and $p_{x}\left(  t\right)  $ for $\beta=0.3$ with the
initial conditions $E=53.0$, $x\left(  0\right)  =2.0$, $p_{x}\left(
0\right)  =0.1$ and $y\left(  0\right)  =0$. The red line and green dot
represent the event horizon and the staring point of the orbit, respectively.
The arrows show the increase in simulation time. The inset displays that the
particle crosses the horizon at $t\simeq0.18748$.}}%
\label{fig:TraE53beta3}%
\end{figure}

With a proper initial condition, we find that a particle of energy $E>E_{\max
}$ can asymptotically approach the event horizon when $\beta=0$ or cross the
horizon within a finite interval of time when $\beta>0$. In fact, the
Hamilton-Jacobi equation $\left(  \ref{eq:HJE}\right)  $ gives%
\begin{equation}
p^{2}\left(  1+\beta p^{2}\right)  ^{2}\sim x^{-2}\text{ as }x\rightarrow0,
\end{equation}
which means $p^{2}=p_{x}^{2}+p_{y}^{2}\rightarrow+\infty$ as $x\rightarrow0$.
On the other hand, if one turns off the harmonic potential $V\left(
x,y,z\right)  $, $p_{y}$ becomes a conserved quantity and always stays finite.
After $V\left(  x,y,z\right)  $ is turned on, it is naturally expected that
$p_{y}$ keeps finite at the horizon even although it is not conserved anymore.
So $p_{x}\rightarrow-\infty$ as $x\rightarrow0$, where we choose $-\infty$
since only ingoing solutions are physical. Near the horizon at $x=0$, the
equations of motion for $x$ and $p_{x}$ then become%
\begin{equation}
\dot{x}\simeq-\alpha x\left(  1+3\beta p_{x}^{2}\right)  \text{ and }\dot
{p}_{x}\simeq\alpha p_{x}\left(  1+\beta p_{x}^{2}\right)  . \label{eq:nhEOM}%
\end{equation}
The solutions to the above equations are%
\begin{equation}
x\simeq A\alpha^{-1}e^{-\alpha\left(  t-t_{0}\right)  }\left[  1-e^{2\alpha
\left(  t-t_{0}\right)  }\alpha^{-2}\beta\right]  ^{3/2}\text{ and }%
p_{x}\simeq-\frac{\alpha^{-1}e^{\alpha\left(  t-t_{0}\right)  }}%
{\sqrt{1-e^{2\alpha\left(  t-t_{0}\right)  }\alpha^{-2}\beta}},
\label{eqn.xpxx0}%
\end{equation}
where $A$ and $t_{0}$ are constants of integration with $e^{2\alpha t_{0}%
}>\alpha^{-2}\beta$. If $\beta=0$, it shows that the particle gets
infinitesimally close to the horizon but never crosses it, which was well
known a long time ago. To better illustrate the $\beta=0$ solutions, we plot
the orbit in the $x$-$y$ plane, $x\left(  t\right)  $ and $p_{x}\left(
t\right)  $ with $E=53.0$, $x\left(  0\right)  =2.0$, $p_{x}\left(  0\right)
=0.1$, $y\left(  0\right)  =0$ and $\beta=0$ in FIG. \ref{fig:TraE53beta0},
which shows that $x$ and $p_{x}$ asymptotically approach $0$ and $-\infty$,
respectively. More interestingly, when $\beta>0$, eqn. $\left(
\ref{eqn.xpxx0}\right)  $ gives that the particle crosses the horizon at
$t=t_{c}\equiv t_{0}-\ln\left(  \beta\alpha^{-2}\right)  /\left(
2\alpha\right)  $, and travels behind the horizon when $t>t_{c}$.\ FIG.
\ref{fig:TraE53beta3} presents the orbit in the $x$-$y$ plane, $x\left(
t\right)  $ and $p_{x}\left(  t\right)  $ with $E=53.0$, $x\left(  0\right)
=2.0$, $p_{x}\left(  0\right)  =0.1$, $y\left(  0\right)  =0$ and $\beta=0.3$.
As shown in the inset, $x\left(  t\right)  =0$ occurs at $t\simeq0.18748$,
where numerical failure is encountered. Note that our near-horizon analysis is
quite universal, regardless of the form of the potential, since no
contributions from the potential appear in eqn. $\left(  \ref{eq:nhEOM}%
\right)  $. 

\subsection{Lyapunov Characteristic Exponents}

\ LCEs have been proposed to describe the time evolution of perturbations of
dynamical systems based on the linearization of equations of motion
\cite{lyapunov1992general}. For a dynamical system that satisfies the
evolution equation $\mathbf{\dot{x}}=f\left(  \mathbf{x}\right)  $, the
evolution of the tangent vectors $\mathbf{Y}$ along a trajectory
$\mathbf{x}\left(  t\right)  $ is determined by%
\begin{equation}
\mathbf{\dot{Y}}=\mathbf{JY},
\end{equation}
where $\mathbf{J}$ is the Jacobian matrix, and $\mathbf{Y}\left(  0\right)
=I$. The matrix $\mathbf{Y}$ characterizes how perturbations of $\mathbf{x}%
\left(  0\right)  $ propagate to the final point $\mathbf{x}\left(  t\right)
$, and defines another matrix $\mathbf{\Lambda}$ in the infinite time limit,%
\begin{equation}
\mathbf{\Lambda}=\lim_{t\rightarrow\infty}\frac{1}{2t}\ln\left[
\mathbf{Y}\left(  t\right)  \mathbf{Y}^{T}\left(  t\right)  \right]  .
\end{equation}
The eigenvalues of $\mathbf{\Lambda}$ are defined as LCEs $\lambda_{i}$, which
measure the exponential expansion rates of infinitesimal perturbations along
the trajectory $\mathbf{x}\left(  t\right)  $. The sum of first $p$ largest
LCEs can be obtained by computing the expansion rate of a $p$-dimensional
volume along $\mathbf{x}\left(  t\right)  $. In practice, one starts with $p$
linearly independent perturbations, evolves them along $\mathbf{x}\left(
t\right)  $ and performs the QR decomposition \cite{geist1990comparison} (or,
equivalently the Gram-Schmidt orthonormalization \cite{benettin1980lyapunov})
at each step to counterbalance all vectors tending to align along the same
direction. The expansion rates are then averaged over $N$ successive steps,
yielding the LCEs spectrum. We here employ the method based on the QR
decomposition to numerically calculate the LCEs $\lambda_{i}$.

Among all LCEs $\lambda_{i}$, the maximum Lyapunov characteristic exponent
(MLCE) $\lambda_{\max}$ is of particularly interesting since a strictly
positive MLCE can be considered as an indication of deterministic chaos. The
MLCE of the motion of a particle near the horizon of the most general static
black hole has recently been argued to satisfy a universal bound
\cite{Hashimoto:2016dfz,Dalui:2018qqv},%
\begin{equation}
\lambda_{\max}\leq\frac{2\pi T}{\hbar}=\alpha, \label{eq:ubound}%
\end{equation}
where $T$ is the temperature, and $\alpha$ is the surface gravity. The bound
$\left(  \ref{eq:ubound}\right)  $ was also conjectured to be satisfied for
MLCEs of out-of-time-ordered correlators in thermal quantum field theories
\cite{Maldacena:2015waa}. Interestingly, it was later shown that the bound
$\left(  \ref{eq:ubound}\right)  $ can be violated for the motion of a charged
massive particle in some charged black hole \cite{Zhao:2018wkl} or when the
minimal length effects are taken into account \cite{Lu:2018mpr}.

\begin{figure}[tb]
\begin{center}
\includegraphics[width=0.45\textwidth]{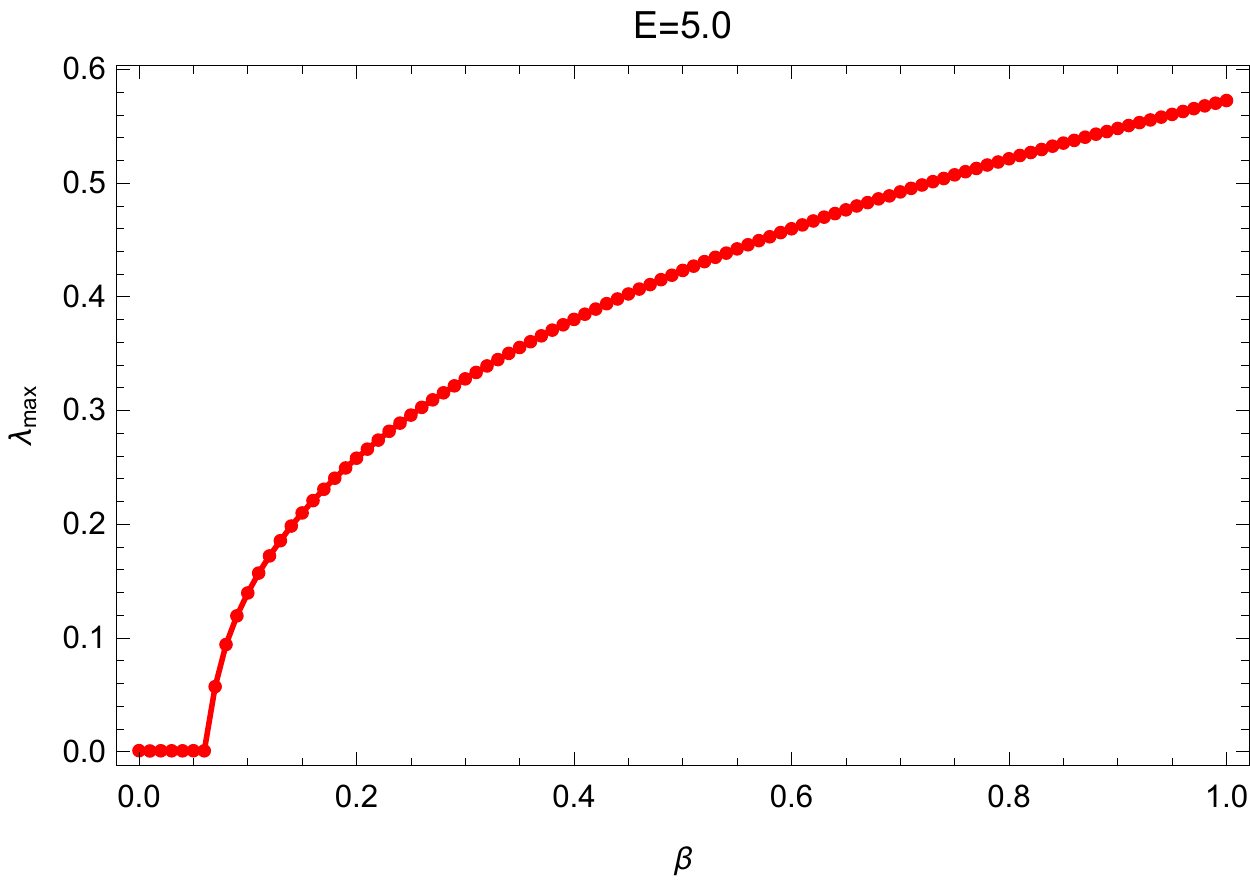}
\includegraphics[width=0.45\textwidth]{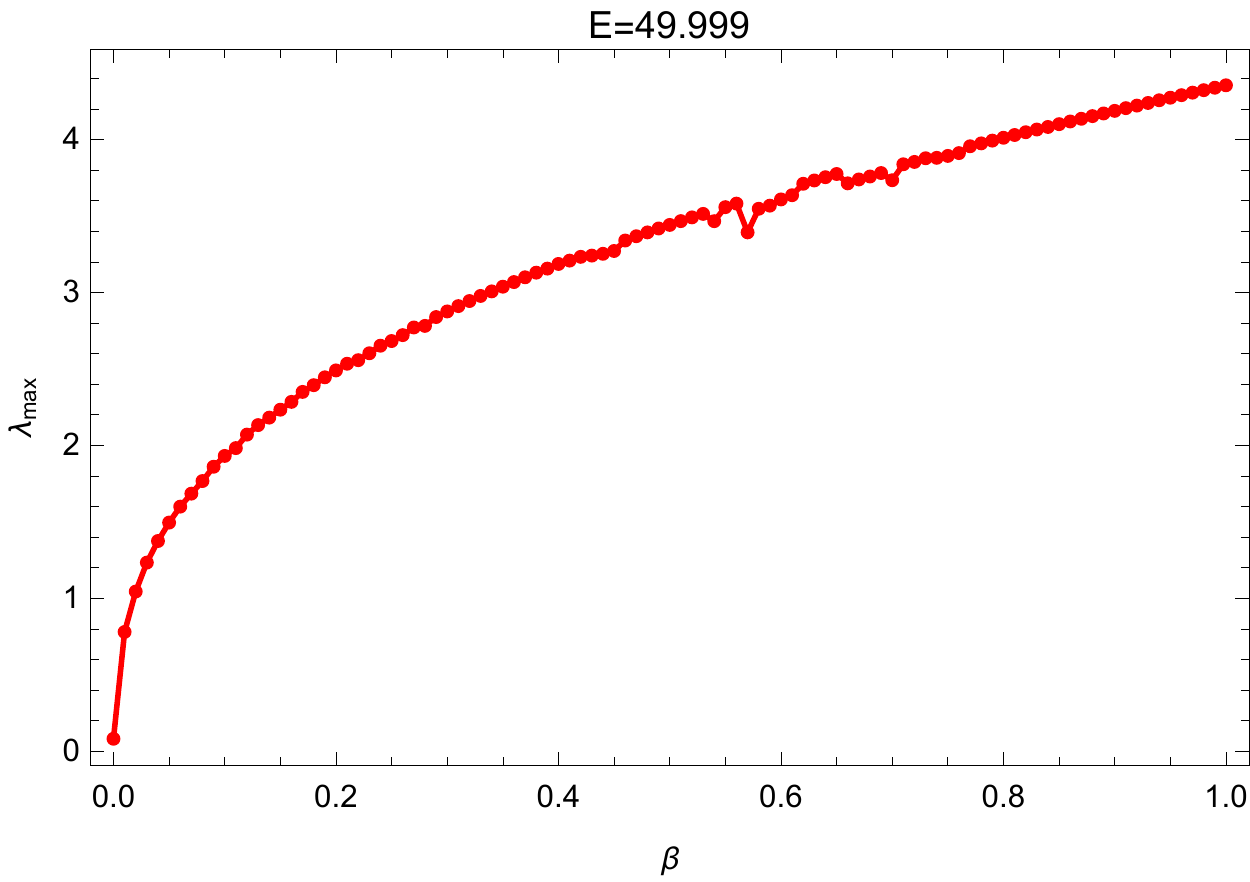}
\end{center}
\caption{{\footnotesize The maximum Lyapunov characteristic exponents
$\lambda_{\text{max}}$ as functions of $\beta$ for $E=5.0$ (\textbf{Left
Panel}) and $49.999$ (\textbf{Right Panel}). It shows that $\lambda
_{\text{max}}$ is always positive, and primarily increases as $\beta$
increases, revealing the minimal length effects could make the trajectories
more chaotic.}}%
\label{fig:MLCE}%
\end{figure}

\ In our case, the dependence of MLCEs on $\beta$ are exhibited in FIG.
\ref{fig:MLCE} for the initial conditions $E=5.0$, $x\left(  0\right)
=0.9999$, $y\left(  0\right)  =0$ and $p_{y}\left(  0\right)  =0$, and
$E=49.999$, $x\left(  0\right)  =1.2$, $y\left(  0\right)  =0$ and
$p_{y}\left(  0\right)  =0$, respectively. Our numerical results show that
MLCEs depend strongly on the choice of the energy $E$, and are quite
insensitive to the remaining initial conditions. For $E=5.0$, we notice that
the MLCE is always positive. As $\beta$ ranges from $0$ to $0.006$, the MLCE
is around $7\times10^{-4}$. Interestingly, the MLCE as a function of $\beta$
has a kink at $\beta\simeq0.006$. When $\beta\gtrsim0.006$, $\beta$ becomes
significantly greater than zero, and grows rapidly as $\beta$ increases. Since
we here choose $\alpha=1$, the MLCE always satisfies the bound $\left(
\ref{eq:ubound}\right)  $. When $E=49.999$, the MLCE remains positive as well,
and has an increasing trend as a function of $\beta$ except several small
fluctuations. It is noteworthy that when $\beta\gtrsim0.02$, the bound
$\left(  \ref{eq:ubound}\right)  $ is violated for the MLCE. As the minimal
length corrections in $\mathcal{H}$ are the order of $\beta E^{2}$, the
minimal length effects play a much more important role in the $E=49.999$ case.
The observations in FIG. \ref{fig:MLCE} indicate that the minimal length
effects tend to make the dynamical system more chaotic, especially when the
energy of the system is large.

\subsection{Poincar\'{e} Surface of Section}

Poincar\'{e} maps were introduced to map complicated behavior in the phase
space to a certain lower-dimensional subspace, called the Poincar\'{e} surface
of section \cite{poincare1881memoire}. Poincar\'{e} maps can be used to
visualize the dynamics of a chaotic system. In an integrable system, a
quasi-periodic orbit fills the Kolmogorov--Arnold--Moser (KAM) torus densely
in the course of time, while a periodic (resonant) orbit repeats itself after
a few windings. On a Poincar\'{e} surface of section intersecting
transversally the torus, the crossing points constitute a closed curve and a
finite number of fixed points for the quasi-periodic and periodic orbits,
respectively. When the integrable system gets perturbed slightly, the KAM
theorem \cite{tabor1989chaos} gives that the quasiperiodic KAM tori are
usually deformed but not destroyed, which also leads to KAM closed curves on
Poincar\'{e} surfaces of section. However, the resonant orbits disintegrate to
form Birkhoff chains of islands and thin chaotic layers surrounding the
Birkhoff islands of stability on Poincar\'{e} surfaces of section
\cite{contopoulos2004order}. On further deviating from the integrable system,
the KAM curves intervening between chaotic layers of different resonances are
destroyed, and the chaotic layers can overlap, which generates large scale
chaos and stronger chaotic behavior. In a nutshell, the observation of a
region with scattered points in a Poincar\'{e} surface of section is a clear
signature of chaos. \begin{figure}[tb]
\begin{center}
\includegraphics[width=0.49\textwidth]{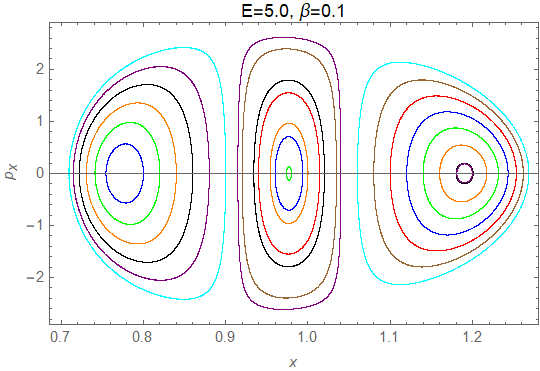}
\includegraphics[width=0.49\textwidth]{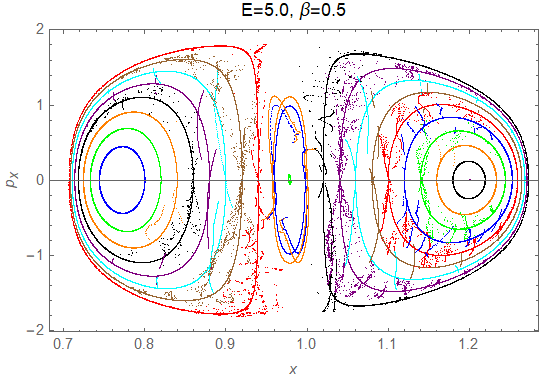}
\includegraphics[width=0.49\textwidth]{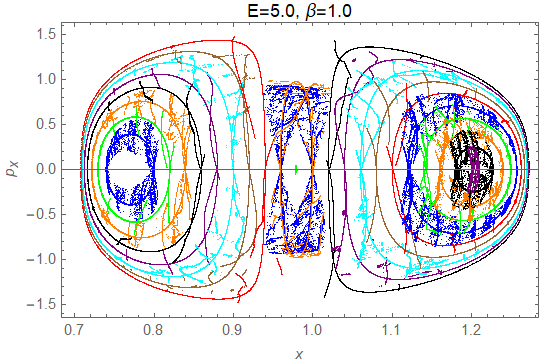}
\includegraphics[width=0.49\textwidth]{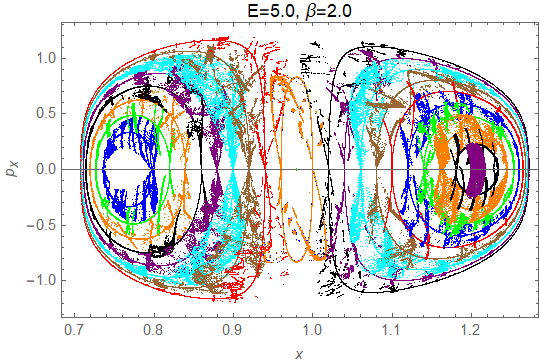}
\end{center}
\caption{{\footnotesize The dependence of the Poincar\'{e} surfaces of section
on $\beta$ for the motion of a particle with $E=5.0$. As $\beta$ increases,
the KAM tori tend to break, which implies that the chaotic behavior becomes
stronger. }}%
\label{fig:PSE5}%
\end{figure}

Here, we choose the Poincar\'{e} surface of section as the one defined by
$y=0$ and $p_{y}>0$, which consequently leaves us a two-dimensional diagram
with the vertical $p_{x}$ axis and the horizontal $x$ axis. FIG.
\ref{fig:PSE5} shows the Poincar\'{e} surfaces of section for $E=5.0$ with
various values of $\beta$. We select 22 initial conditions (i.e., $E$,
$x\left(  0\right)  $, $p_{x}\left(  0\right)  $ and $y\left(  0\right)  $),
and plot the crossing points of the corresponding phase orbits in 8 colors on
the Poincar\'{e} surfaces of section. The crossing points of the same orbit
are in the same color, whereas one color corresponds to $2$ or $3$ orbits.
Note that the set of initial conditions of the orbits is the same for each
Poincar\'{e} surface of section. When $\beta=0.1$, the Poincar\'{e} surface of
section consists of KAM closed curves, which shows that the orbits are
quasi-periodic. This observation is in agreement with FIG. \ref{fig:TraE5},
which also reveals that the orbit with small $\beta$ is quasi-periodic. But
for a larger $\beta$, it displays that the KAM curves start to get destroyed
to become scattered plots (e.g., see the lower-right panel in FIG.
\ref{fig:PSE5}). In short, the growth of the minimal length effects leads to
the onset of chaos and a following increase in chaotic behavior of the motion
of a particle in the Rindler space.

\begin{figure}[tb]
\begin{center}
\includegraphics[width=0.49\textwidth]{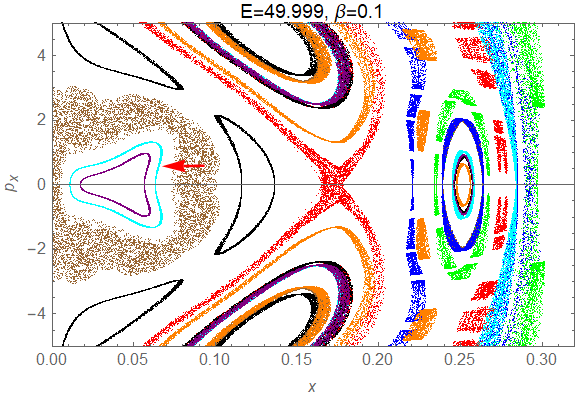}
\includegraphics[width=0.49\textwidth]{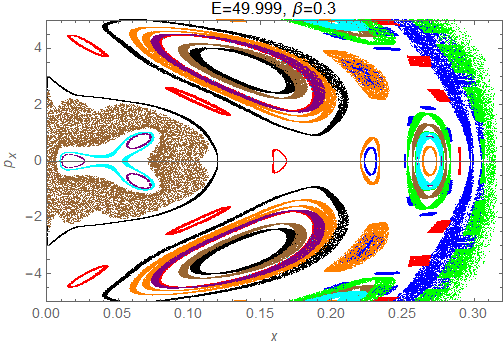}
\includegraphics[width=0.49\textwidth]{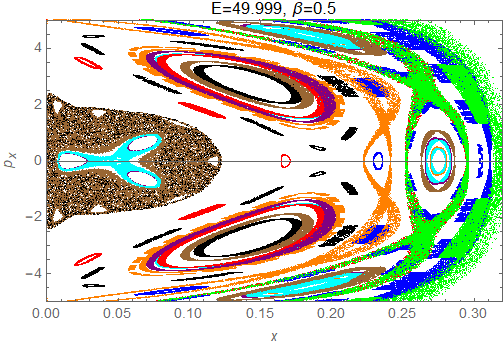}
\includegraphics[width=0.49\textwidth]{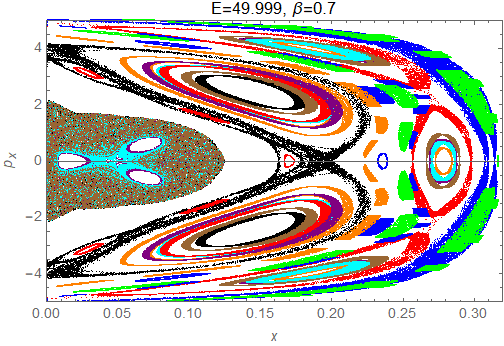}
\end{center}
\caption{{\footnotesize The dependence of the Poincar\'{e} surfaces of section
on $\beta$ for the motion of a particle with $E=49.999$. As $\beta$ increases
from $0.1$ to $0.7$, the cyan KAM curve marked by a red arrow first gets
deformed ($\beta=0.3$), and then disintegrates to form a chaotic region
($\beta=0.5$), which becomes more dispersed ($\beta=0.7$). So as the minimal
length effects are increased, the chaotic features become more evident. }}%
\label{fig:PSE49}%
\end{figure}

When the energy increases, chaotic regions in Poincar\'{e} surfaces of section
can be observed for a small value of $\beta$ or even $\beta=0$. We present the
dependence of the Poincar\'{e} surfaces of section on $\beta$ with $E=$
$49.999$ in FIG. \ref{fig:PSE49}, which exhibits much richer structures than
the $E=$ $5.0$ case. In fact, chains of islands, chaotic regions with
scattered plots and separatrices in the Poincar\'{e} surfaces of section and
their evolution with $\beta$ are seen in FIG. \ref{fig:PSE49}, where we plot
crossing points of 31 phase orbits in 8 colors. To study the minimal length
effects on the chaotic behavior, we focus on the cyan KAM curve on the left
side of the $\beta=0$ Poincar\'{e} surface of section (the upper-left panel of
FIG. \ref{fig:PSE49}), which is marked by a red arrow and encircled by the
chaotic region with scattered brown points. For $\beta=0.3$, it shows that the
cyan KAM curve is not destroyed but gets deformed with a larger width. When
$\beta$ is increased to as large as $0.5$, the lower-left panel of FIG.
\ref{fig:PSE49} displays that the cyan KAM curve disintegrates into scattered
cyan points so as to form a chaotic region. On further increasing $\beta$ from
$0.5$ to $0.7$, the chaotic region becomes more dispersed on the Poincar\'{e}
surface of section, which is shown in the lower-right panel of FIG.
\ref{fig:PSE49}. These observations lend further support to the conclusion
that the minimal length effects can make chaotic behavior of the system stronger.

\section{Discussion and Conclusion}

\label{Sec:DC}

In this paper, we investigated the minimal length effects on the motion of a
particle in the Rindler space under a harmonic potential. We first
distinguished two different types of trajectories in the motion of the
particle. For the first type of trajectories, the particle travels around the
fixed point, and the trajectories can be erratic when the minimal length
effects are large enough. The particle moving along the second type of
trajectories will cross the horizon at a finite Rindler time if the minimal
length effects are turned on, whereas it just asymptotically approaches the
event horizon in the absence of the minimal length effects. We then exploited
Poincar\'{e} surfaces of section and LCEs to investigate the chaotic behavior
of the system and found that, as the minimal length effects grow,

\begin{itemize}
\item FIG. \ref{fig:MLCE} showed that, for $E=5.0$ and $49.999$, the MLCEs are
always positive and generally increase.

\item FIG. \ref{fig:PSE5} displayed that, for $E=5.0$, the KAM curves tend to disintegrate.

\item FIG. \ref{fig:PSE49} exhibited that, for $E=49.999$, the cyan KAM curve
breaks into a chaotic layer.
\end{itemize}

In light of our numerical results, we come to the conclusion that chaotic
behavior is more likely to happen in the presence of the minimal length
effects. This is in agreement with earlier observations and generic arguments
for a massive particle perturbed away from an unstable equilibrium near the
black hole horizon \cite{Lu:2018mpr}, as well as recent findings for the
geodesic motion perturbed by the minimal length effects around a Schwarzschild
black hole \cite{Guo:2020xnf}. In addition, black hole horizons have been
conjectured to be fastest scramblers in nature \cite{Sekino:2008he} with the
scrambling time $t_{s}\sim\hbar T^{-1}\ln S$, where $T$ and $S$ are the
temperature and entropy of the black hole, respectively. Here, we can use eqn.
$\left(  \ref{eqn.xpxx0}\right)  $ to estimate $t_{s}$ by relating $t_{s}$ to
the time that it takes to reach the stretched horizon located at $x=\delta$,
which is roughly one Planck length $\ell_{p}$
\cite{Fischler:2015cma,Guo:2017bru}. Then eqns. $\left(  \ref{eq:ht}\right)  $
and $\left(  \ref{eqn.xpxx0}\right)  $ give%
\begin{equation}
t_{s}\sim\frac{\hbar}{2\pi T}\left[  \ln\frac{\ell_{p}}{2\pi T}-\frac
{3\beta_{0}}{2}\left(  \frac{\ell_{p}}{2\pi T}\right)  ^{4}\right]  ,
\end{equation}
where we define a dimensionless parameter $\beta_{0}\equiv\beta\ell_{p}^{2}$.
For a Schwarzschild black hole, the scrambling time $t_{s}$ is%
\begin{equation}
t_{s}\sim\frac{\hbar}{2\pi T}\left(  \ln\frac{16S}{\pi^{2}}-\frac{3\beta
_{0}\pi^{4}}{512S^{2}}\right)  ,
\end{equation}
which indicates that the minimal length effects can make black holes scramble
faster. To summarize, in this paper we proposed a toy model to show that
quantum gravity effects tend to increase chaotic behavior and scrambling
efficiency of black holes. Further exploration of quantum gravity effects on
chaotic dynamics will lend insight into physics of black holes, early universe
and dynamical astronomy.

\begin{acknowledgments}
We are grateful to Houwen Wu and Haitang Yang for useful discussions. This
work is supported in part by NSFC (Grant No. 11875196, 11375121 and 11005016),
the Fundamental Research Funds for the Central Universities, Natural Science
Foundation of Chengdu University of TCM (Grants nos. ZRYY1729 and ZRYY1921),
Discipline Talent Promotion Program of /Xinglin Scholars(Grant no.
QNXZ2018050), the key fund project for Education Department of Sichuan (Grant
no. 18ZA0173) and Special Talent Projects of Chizhou University (Grant no. RZ2000000591).
\end{acknowledgments}


\begin{thebibliography}{10}

\bibitem{Barrow:1981sx}
John~D. Barrow.
\newblock {Chaotic behavior in general relativity}.
\newblock {\em Phys. Rept.}, 85:1--49, 1982.
\newblock \href {https://doi.org/10.1016/0370-1573(82)90171-5}
  {\path{doi:10.1016/0370-1573(82)90171-5}}.

\bibitem{Motter:2000bg}
A.E. Motter and P.S. Letelier.
\newblock {Mixmaster chaos}.
\newblock {\em Phys. Lett. A}, 285:127--131, 2001.
\newblock \href {http://arxiv.org/abs/gr-qc/0011001}
  {\path{arXiv:gr-qc/0011001}}, \href
  {https://doi.org/10.1016/S0375-9601(01)00349-8}
  {\path{doi:10.1016/S0375-9601(01)00349-8}}.

\bibitem{Carter:1968rr}
Brandon Carter.
\newblock {Global structure of the Kerr family of gravitational fields}.
\newblock {\em Phys. Rev.}, 174:1559--1571, 1968.
\newblock \href {https://doi.org/10.1103/PhysRev.174.1559}
  {\path{doi:10.1103/PhysRev.174.1559}}.

\bibitem{Sota:1995ms}
Yasuhide Sota, Shingo Suzuki, and Kei-ichi Maeda.
\newblock {Chaos in static axisymmetric space-times. 1: Vacuum case}.
\newblock {\em Class. Quant. Grav.}, 13:1241--1260, 1996.
\newblock \href {http://arxiv.org/abs/gr-qc/9505036}
  {\path{arXiv:gr-qc/9505036}}, \href
  {https://doi.org/10.1088/0264-9381/13/5/034}
  {\path{doi:10.1088/0264-9381/13/5/034}}.

\bibitem{Hanan:2006uf}
William Hanan and Eugen Radu.
\newblock {Chaotic motion in multi-black hole spacetimes and holographic
  screens}.
\newblock {\em Mod. Phys. Lett.}, A22:399--406, 2007.
\newblock \href {http://arxiv.org/abs/gr-qc/0610119}
  {\path{arXiv:gr-qc/0610119}}, \href
  {https://doi.org/10.1142/S0217732307022815}
  {\path{doi:10.1142/S0217732307022815}}.

\bibitem{Gair:2007kr}
Jonathan~R. Gair, Chao Li, and Ilya Mandel.
\newblock {Observable Properties of Orbits in Exact Bumpy Spacetimes}.
\newblock {\em Phys. Rev.}, D77:024035, 2008.
\newblock \href {http://arxiv.org/abs/0708.0628} {\path{arXiv:0708.0628}},
  \href {https://doi.org/10.1103/PhysRevD.77.024035}
  {\path{doi:10.1103/PhysRevD.77.024035}}.

\bibitem{Zahrani:2013up}
A.M.Al Zahrani, Valeri~P. Frolov, and Andrey~A. Shoom.
\newblock {Critical escape velocity for a charged particle moving around a
  weakly magnetized Schwarzschild black hole}.
\newblock {\em Phys. Rev. D}, 87(8):084043, 2013.
\newblock \href {http://arxiv.org/abs/1301.4633} {\path{arXiv:1301.4633}},
  \href {https://doi.org/10.1103/PhysRevD.87.084043}
  {\path{doi:10.1103/PhysRevD.87.084043}}.

\bibitem{Witzany:2015yqa}
V.~Witzany, O.~Semerák, and P.~Suková.
\newblock {Free motion around black holes with discs or rings: between
  integrability and chaos – IV}.
\newblock {\em Mon. Not. Roy. Astron. Soc.}, 451(2):1770--1794, 2015.
\newblock \href {http://arxiv.org/abs/1503.09077} {\path{arXiv:1503.09077}},
  \href {https://doi.org/10.1093/mnras/stv1148}
  {\path{doi:10.1093/mnras/stv1148}}.

\bibitem{Wang:2016wcj}
Mingzhi Wang, Songbai Chen, and Jiliang Jing.
\newblock {Chaos in the motion of a test scalar particle coupling to the
  Einstein tensor in Schwarzschild–Melvin black hole spacetime}.
\newblock {\em Eur. Phys. J.}, C77(4):208, 2017.
\newblock \href {http://arxiv.org/abs/1605.09506} {\path{arXiv:1605.09506}},
  \href {https://doi.org/10.1140/epjc/s10052-017-4792-y}
  {\path{doi:10.1140/epjc/s10052-017-4792-y}}.

\bibitem{Chen:2016tmr}
Songbai Chen, Mingzhi Wang, and Jiliang Jing.
\newblock {Chaotic motion of particles in the accelerating and rotating black
  holes spacetime}.
\newblock {\em JHEP}, 09:082, 2016.
\newblock \href {http://arxiv.org/abs/1604.02785} {\path{arXiv:1604.02785}},
  \href {https://doi.org/10.1007/JHEP09(2016)082}
  {\path{doi:10.1007/JHEP09(2016)082}}.

\bibitem{Wang:2018eui}
Mingzhi Wang, Songbai Chen, and Jiliang Jing.
\newblock {Chaotic shadow of a non-Kerr rotating compact object with quadrupole
  mass moment}.
\newblock {\em Phys. Rev. D}, 98(10):104040, 2018.
\newblock \href {http://arxiv.org/abs/1801.02118} {\path{arXiv:1801.02118}},
  \href {https://doi.org/10.1103/PhysRevD.98.104040}
  {\path{doi:10.1103/PhysRevD.98.104040}}.

\bibitem{Hashimoto:2016dfz}
Koji Hashimoto and Norihiro Tanahashi.
\newblock {Universality in Chaos of Particle Motion near Black Hole Horizon}.
\newblock {\em Phys. Rev.}, D95(2):024007, 2017.
\newblock \href {http://arxiv.org/abs/1610.06070} {\path{arXiv:1610.06070}},
  \href {https://doi.org/10.1103/PhysRevD.95.024007}
  {\path{doi:10.1103/PhysRevD.95.024007}}.

\bibitem{Dalui:2019umw}
Surojit Dalui, Bibhas~Ranjan Majhi, and Pankaj Mishra.
\newblock {Induction of chaotic fluctuations in particle dynamics in a
  uniformly accelerated frame}.
\newblock {\em Int. J. Mod. Phys. A}, 35(18):2050081, 2020.
\newblock \href {http://arxiv.org/abs/1904.11760} {\path{arXiv:1904.11760}},
  \href {https://doi.org/10.1142/S0217751X20500815}
  {\path{doi:10.1142/S0217751X20500815}}.

\bibitem{Dalui:2019esx}
Surojit Dalui, Bibhas~Ranjan Majhi, and Pankaj Mishra.
\newblock {Horizon induces instability and creates quantum thermality}.
\newblock 10 2019.
\newblock \href {http://arxiv.org/abs/1910.07989} {\path{arXiv:1910.07989}}.

\bibitem{Apostolatos:2009vu}
Theocharis~A. Apostolatos, Georgios Lukes-Gerakopoulos, and George Contopoulos.
\newblock {How to Observe a Non-Kerr Spacetime Using Gravitational Waves}.
\newblock {\em Phys. Rev. Lett.}, 103:111101, 2009.
\newblock \href {http://arxiv.org/abs/0906.0093} {\path{arXiv:0906.0093}},
  \href {https://doi.org/10.1103/PhysRevLett.103.111101}
  {\path{doi:10.1103/PhysRevLett.103.111101}}.

\bibitem{Gutierrez-Ruiz:2018tre}
Andrés~F. Gutiérrez-Ruiz, Alejandro Cárdenas-Avendaño, Nicolás Yunes, and
  Leonardo~A. Pachón.
\newblock {Stealth Chaos due to Frame Dragging}.
\newblock 6 2018.
\newblock \href {http://arxiv.org/abs/1806.06476} {\path{arXiv:1806.06476}}.

\bibitem{Zayas:2010fs}
Leopoldo~A. Pando~Zayas and Cesar~A. Terrero-Escalante.
\newblock {Chaos in the Gauge / Gravity Correspondence}.
\newblock {\em JHEP}, 09:094, 2010.
\newblock \href {http://arxiv.org/abs/1007.0277} {\path{arXiv:1007.0277}},
  \href {https://doi.org/10.1007/JHEP09(2010)094}
  {\path{doi:10.1007/JHEP09(2010)094}}.

\bibitem{Ma:2014aha}
Da-Zhu Ma, Jian-Pin Wu, and Jifang Zhang.
\newblock {Chaos from the ring string in a Gauss-Bonnet black hole in AdS5
  space}.
\newblock {\em Phys. Rev.}, D89(8):086011, 2014.
\newblock \href {http://arxiv.org/abs/1405.3563} {\path{arXiv:1405.3563}},
  \href {https://doi.org/10.1103/PhysRevD.89.086011}
  {\path{doi:10.1103/PhysRevD.89.086011}}.

\bibitem{Basu:2016zkr}
Pallab Basu, Pankaj Chaturvedi, and Prasant Samantray.
\newblock {Chaotic dynamics of strings in charged black hole backgrounds}.
\newblock {\em Phys. Rev. D}, 95(6):066014, 2017.
\newblock \href {http://arxiv.org/abs/1607.04466} {\path{arXiv:1607.04466}},
  \href {https://doi.org/10.1103/PhysRevD.95.066014}
  {\path{doi:10.1103/PhysRevD.95.066014}}.

\bibitem{Hashimoto:2018fkb}
Koji Hashimoto, Keiju Murata, and Norihiro Tanahashi.
\newblock {Chaos of Wilson Loop from String Motion near Black Hole Horizon}.
\newblock {\em Phys. Rev. D}, 98(8):086007, 2018.
\newblock \href {http://arxiv.org/abs/1803.06756} {\path{arXiv:1803.06756}},
  \href {https://doi.org/10.1103/PhysRevD.98.086007}
  {\path{doi:10.1103/PhysRevD.98.086007}}.

\bibitem{Cubrovic:2019qee}
Mihailo {\v{C}}ubrovi{\'c}.
\newblock {The bound on chaos for closed strings in Anti-de Sitter black hole
  backgrounds}.
\newblock {\em JHEP}, 12:150, 2019.
\newblock \href {http://arxiv.org/abs/1904.06295} {\path{arXiv:1904.06295}},
  \href {https://doi.org/10.1007/JHEP12(2019)150}
  {\path{doi:10.1007/JHEP12(2019)150}}.

\bibitem{Ma:2019ewq}
Da-Zhu Ma, Dan Zhang, Guoyang Fu, and Jian-Pin Wu.
\newblock {Chaotic dynamics of string around charged black brane with
  hyperscaling violation}.
\newblock {\em JHEP}, 01:103, 2020.
\newblock \href {http://arxiv.org/abs/1911.09913} {\path{arXiv:1911.09913}},
  \href {https://doi.org/10.1007/JHEP01(2020)103}
  {\path{doi:10.1007/JHEP01(2020)103}}.

\bibitem{Frolov:1999pj}
Andrei~V. Frolov and Arne~L. Larsen.
\newblock {Chaotic scattering and capture of strings by black hole}.
\newblock {\em Class. Quant. Grav.}, 16:3717--3724, 1999.
\newblock \href {http://arxiv.org/abs/gr-qc/9908039}
  {\path{arXiv:gr-qc/9908039}}, \href
  {https://doi.org/10.1088/0264-9381/16/11/316}
  {\path{doi:10.1088/0264-9381/16/11/316}}.

\bibitem{Suzuki:1996gm}
Shingo Suzuki and Kei-ichi Maeda.
\newblock {Chaos in Schwarzschild space-time: The motion of a spinning
  particle}.
\newblock {\em Phys. Rev. D}, 55:4848--4859, 1997.
\newblock \href {http://arxiv.org/abs/gr-qc/9604020}
  {\path{arXiv:gr-qc/9604020}}, \href
  {https://doi.org/10.1103/PhysRevD.55.4848}
  {\path{doi:10.1103/PhysRevD.55.4848}}.

\bibitem{Hartl:2003da}
Michael~D. Hartl.
\newblock {A Survey of spinning test particle orbits in Kerr space-time}.
\newblock {\em Phys. Rev. D}, 67:104023, 2003.
\newblock \href {http://arxiv.org/abs/gr-qc/0302103}
  {\path{arXiv:gr-qc/0302103}}, \href
  {https://doi.org/10.1103/PhysRevD.67.104023}
  {\path{doi:10.1103/PhysRevD.67.104023}}.

\bibitem{Lukes-Gerakopoulos:2016bup}
Georgios Lukes-Gerakopoulos, Matthaios Katsanikas, Panos~A. Patsis, and
  Jonathan Seyrich.
\newblock {The dynamics of a spinning particle in a linear in spin Hamiltonian
  approximation}.
\newblock {\em Phys. Rev. D}, 94(2):024024, 2016.
\newblock \href {http://arxiv.org/abs/1606.09171} {\path{arXiv:1606.09171}},
  \href {https://doi.org/10.1103/PhysRevD.94.024024}
  {\path{doi:10.1103/PhysRevD.94.024024}}.

\bibitem{Zelenka:2019nyp}
Ond{\v{r}}ej Zelenka, Georgios Lukes-Gerakopoulos, Vojt{\v{e}}ch Witzany, and
  Ond{\v{r}}ej Kop{\'a}{\v{c}}ek.
\newblock {Growth of resonances and chaos for a spinning test particle in the
  Schwarzschild background}.
\newblock {\em Phys. Rev. D}, 101(2):024037, 2020.
\newblock \href {http://arxiv.org/abs/1911.00414} {\path{arXiv:1911.00414}},
  \href {https://doi.org/10.1103/PhysRevD.101.024037}
  {\path{doi:10.1103/PhysRevD.101.024037}}.

\bibitem{Veneziano:1986zf}
G.~Veneziano.
\newblock {A Stringy Nature Needs Just Two Constants}.
\newblock {\em Europhys. Lett.}, 2:199, 1986.
\newblock \href {https://doi.org/10.1209/0295-5075/2/3/006}
  {\path{doi:10.1209/0295-5075/2/3/006}}.

\bibitem{Gross:1987ar}
David~J. Gross and Paul~F. Mende.
\newblock {String Theory Beyond the Planck Scale}.
\newblock {\em Nucl. Phys.}, B303:407--454, 1988.
\newblock \href {https://doi.org/10.1016/0550-3213(88)90390-2}
  {\path{doi:10.1016/0550-3213(88)90390-2}}.

\bibitem{Amati:1988tn}
D.~Amati, M.~Ciafaloni, and G.~Veneziano.
\newblock {Can Space-Time Be Probed Below the String Size?}
\newblock {\em Phys. Lett.}, B216:41--47, 1989.
\newblock \href {https://doi.org/10.1016/0370-2693(89)91366-X}
  {\path{doi:10.1016/0370-2693(89)91366-X}}.

\bibitem{Garay:1994en}
Luis~J. Garay.
\newblock {Quantum gravity and minimum length}.
\newblock {\em Int. J. Mod. Phys.}, A10:145--166, 1995.
\newblock \href {http://arxiv.org/abs/gr-qc/9403008}
  {\path{arXiv:gr-qc/9403008}}, \href
  {https://doi.org/10.1142/S0217751X95000085}
  {\path{doi:10.1142/S0217751X95000085}}.

\bibitem{Scardigli:1999jh}
Fabio Scardigli.
\newblock {Generalized uncertainty principle in quantum gravity from micro -
  black hole Gedanken experiment}.
\newblock {\em Phys. Lett. B}, 452:39--44, 1999.
\newblock \href {http://arxiv.org/abs/hep-th/9904025}
  {\path{arXiv:hep-th/9904025}}, \href
  {https://doi.org/10.1016/S0370-2693(99)00167-7}
  {\path{doi:10.1016/S0370-2693(99)00167-7}}.

\bibitem{Maggiore:1993kv}
Michele Maggiore.
\newblock {The Algebraic structure of the generalized uncertainty principle}.
\newblock {\em Phys. Lett.}, B319:83--86, 1993.
\newblock \href {http://arxiv.org/abs/hep-th/9309034}
  {\path{arXiv:hep-th/9309034}}, \href
  {https://doi.org/10.1016/0370-2693(93)90785-G}
  {\path{doi:10.1016/0370-2693(93)90785-G}}.

\bibitem{Kempf:1994su}
Achim Kempf, Gianpiero Mangano, and Robert~B. Mann.
\newblock {Hilbert space representation of the minimal length uncertainty
  relation}.
\newblock {\em Phys. Rev.}, D52:1108--1118, 1995.
\newblock \href {http://arxiv.org/abs/hep-th/9412167}
  {\path{arXiv:hep-th/9412167}}, \href
  {https://doi.org/10.1103/PhysRevD.52.1108}
  {\path{doi:10.1103/PhysRevD.52.1108}}.

\bibitem{Chang:2001kn}
Lay~Nam Chang, Djordje Minic, Naotoshi Okamura, and Tatsu Takeuchi.
\newblock {Exact solution of the harmonic oscillator in arbitrary dimensions
  with minimal length uncertainty relations}.
\newblock {\em Phys. Rev.}, D65:125027, 2002.
\newblock \href {http://arxiv.org/abs/hep-th/0111181}
  {\path{arXiv:hep-th/0111181}}, \href
  {https://doi.org/10.1103/PhysRevD.65.125027}
  {\path{doi:10.1103/PhysRevD.65.125027}}.

\bibitem{Akhoury:2003kc}
R.~Akhoury and Y.~P. Yao.
\newblock {Minimal length uncertainty relation and the hydrogen spectrum}.
\newblock {\em Phys. Lett.}, B572:37--42, 2003.
\newblock \href {http://arxiv.org/abs/hep-ph/0302108}
  {\path{arXiv:hep-ph/0302108}}, \href
  {https://doi.org/10.1016/j.physletb.2003.07.084}
  {\path{doi:10.1016/j.physletb.2003.07.084}}.

\bibitem{Brau:1999uv}
F.~Brau.
\newblock {Minimal length uncertainty relation and hydrogen atom}.
\newblock {\em J. Phys.}, A32:7691--7696, 1999.
\newblock \href {http://arxiv.org/abs/quant-ph/9905033}
  {\path{arXiv:quant-ph/9905033}}, \href
  {https://doi.org/10.1088/0305-4470/32/44/308}
  {\path{doi:10.1088/0305-4470/32/44/308}}.

\bibitem{Brau:2006ca}
Fabian Brau and Fabien Buisseret.
\newblock {Minimal Length Uncertainty Relation and gravitational quantum well}.
\newblock {\em Phys. Rev.}, D74:036002, 2006.
\newblock \href {http://arxiv.org/abs/hep-th/0605183}
  {\path{arXiv:hep-th/0605183}}, \href
  {https://doi.org/10.1103/PhysRevD.74.036002}
  {\path{doi:10.1103/PhysRevD.74.036002}}.

\bibitem{Pedram:2011xj}
Pouria Pedram, Kourosh Nozari, and S.~H. Taheri.
\newblock {The effects of minimal length and maximal momentum on the transition
  rate of ultra cold neutrons in gravitational field}.
\newblock {\em JHEP}, 03:093, 2011.
\newblock \href {http://arxiv.org/abs/1103.1015} {\path{arXiv:1103.1015}},
  \href {https://doi.org/10.1007/JHEP03(2011)093}
  {\path{doi:10.1007/JHEP03(2011)093}}.

\bibitem{Pikovski:2011zk}
Igor Pikovski, Michael~R. Vanner, Markus Aspelmeyer, M.~S. Kim, and Caslav
  Brukner.
\newblock {Probing Planck-scale physics with quantum optics}.
\newblock {\em Nature Phys.}, 8:393--397, 2012.
\newblock \href {http://arxiv.org/abs/1111.1979} {\path{arXiv:1111.1979}},
  \href {https://doi.org/10.1038/nphys2262} {\path{doi:10.1038/nphys2262}}.

\bibitem{Bosso:2018ckz}
Pasquale Bosso, Saurya Das, and Robert~B. Mann.
\newblock {Potential tests of the Generalized Uncertainty Principle in the
  advanced LIGO experiment}.
\newblock {\em Phys. Lett.}, B785:498--505, 2018.
\newblock \href {http://arxiv.org/abs/1804.03620} {\path{arXiv:1804.03620}},
  \href {https://doi.org/10.1016/j.physletb.2018.08.061}
  {\path{doi:10.1016/j.physletb.2018.08.061}}.

\bibitem{Wang:2010ct}
Peng Wang, Haitang Yang, and Xiuming Zhang.
\newblock {Quantum gravity effects on statistics and compact star
  configurations}.
\newblock {\em JHEP}, 08:043, 2010.
\newblock \href {http://arxiv.org/abs/1006.5362} {\path{arXiv:1006.5362}},
  \href {https://doi.org/10.1007/JHEP08(2010)043}
  {\path{doi:10.1007/JHEP08(2010)043}}.

\bibitem{Wang:2011iv}
Peng Wang, Haitang Yang, and Xiuming Zhang.
\newblock {Quantum gravity effects on compact star cores}.
\newblock {\em Phys. Lett. B}, 718:265--269, 2012.
\newblock \href {http://arxiv.org/abs/1110.5550} {\path{arXiv:1110.5550}},
  \href {https://doi.org/10.1016/j.physletb.2012.10.071}
  {\path{doi:10.1016/j.physletb.2012.10.071}}.

\bibitem{Ong:2018zqn}
Yen~Chin Ong.
\newblock {Generalized Uncertainty Principle, Black Holes, and White Dwarfs: A
  Tale of Two Infinities}.
\newblock {\em JCAP}, 1809:015, 2018.
\newblock \href {http://arxiv.org/abs/1804.05176} {\path{arXiv:1804.05176}},
  \href {https://doi.org/10.1088/1475-7516/2018/09/015}
  {\path{doi:10.1088/1475-7516/2018/09/015}}.

\bibitem{Guo:2016btf}
Xiaobo Guo, Bochen Lv, Jun Tao, and Peng Wang.
\newblock {Quantum Tunneling In Deformed Quantum Mechanics with Minimal
  Length}.
\newblock {\em Adv. High Energy Phys.}, 2016:4502312, 2016.
\newblock \href {http://arxiv.org/abs/1609.06944} {\path{arXiv:1609.06944}},
  \href {https://doi.org/10.1155/2016/4502312}
  {\path{doi:10.1155/2016/4502312}}.

\bibitem{Khodadi:2018wed}
Mohsen Khodadi, Kourosh Nozari, and Fazlollah Hajkarim.
\newblock {On the viability of Planck scale cosmology with quartessence}.
\newblock {\em Eur. Phys. J. C}, 78(9):716, 2018.
\newblock \href {http://arxiv.org/abs/1808.08436} {\path{arXiv:1808.08436}},
  \href {https://doi.org/10.1140/epjc/s10052-018-6191-4}
  {\path{doi:10.1140/epjc/s10052-018-6191-4}}.

\bibitem{Khodadi:2018scn}
Mohsen Khodadi, Kourosh Nozari, Habib Abedi, and Salvatore Capozziello.
\newblock {Planck scale effects on the stochastic gravitational wave background
  generated from cosmological hadronization transition: A qualitative study}.
\newblock {\em Phys. Lett. B}, 783:326--333, 2018.
\newblock \href {http://arxiv.org/abs/1805.11310} {\path{arXiv:1805.11310}},
  \href {https://doi.org/10.1016/j.physletb.2018.07.010}
  {\path{doi:10.1016/j.physletb.2018.07.010}}.

\bibitem{Benczik:2002tt}
Sandor Benczik, Lay~Nam Chang, Djordje Minic, Naotoshi Okamura, Saifuddin
  Rayyan, and Tatsu Takeuchi.
\newblock {Short distance versus long distance physics: The Classical limit of
  the minimal length uncertainty relation}.
\newblock {\em Phys. Rev.}, D66:026003, 2002.
\newblock \href {http://arxiv.org/abs/hep-th/0204049}
  {\path{arXiv:hep-th/0204049}}, \href
  {https://doi.org/10.1103/PhysRevD.66.026003}
  {\path{doi:10.1103/PhysRevD.66.026003}}.

\bibitem{Ahmadi:2014cga}
Fatemeh Ahmadi and Jafar Khodagholizadeh.
\newblock {Effect of GUP on the Kepler problem and a variable minimal length}.
\newblock {\em Can. J. Phys.}, 92:484--487, 2014.
\newblock \href {http://arxiv.org/abs/1411.0241} {\path{arXiv:1411.0241}},
  \href {https://doi.org/10.1139/cjp-2013-0354}
  {\path{doi:10.1139/cjp-2013-0354}}.

\bibitem{Silagadze:2009vu}
Zurab~K. Silagadze.
\newblock {Quantum gravity, minimum length and Keplerian orbits}.
\newblock {\em Phys. Lett.}, A373:2643--2645, 2009.
\newblock \href {http://arxiv.org/abs/0901.1258} {\path{arXiv:0901.1258}},
  \href {https://doi.org/10.1016/j.physleta.2009.05.053}
  {\path{doi:10.1016/j.physleta.2009.05.053}}.

\bibitem{Scardigli:2014qka}
Fabio Scardigli and Roberto Casadio.
\newblock {Gravitational tests of the Generalized Uncertainty Principle}.
\newblock {\em Eur. Phys. J. C}, 75(9):425, 2015.
\newblock \href {http://arxiv.org/abs/1407.0113} {\path{arXiv:1407.0113}},
  \href {https://doi.org/10.1140/epjc/s10052-015-3635-y}
  {\path{doi:10.1140/epjc/s10052-015-3635-y}}.

\bibitem{Ali:2015zua}
Ahmed Farag~Ali, Mohammed~M. Khalil, and Elias~C. Vagenas.
\newblock {Minimal Length in quantum gravity and gravitational measurements}.
\newblock {\em EPL}, 112(2):20005, 2015.
\newblock \href {http://arxiv.org/abs/1510.06365} {\path{arXiv:1510.06365}},
  \href {https://doi.org/10.1209/0295-5075/112/20005}
  {\path{doi:10.1209/0295-5075/112/20005}}.

\bibitem{Guo:2015ldd}
Xiaobo Guo, Peng Wang, and Haitang Yang.
\newblock {The classical limit of minimal length uncertainty relation: revisit
  with the Hamilton-Jacobi method}.
\newblock {\em JCAP}, 1605:062, 2016.
\newblock \href {http://arxiv.org/abs/1512.03560} {\path{arXiv:1512.03560}},
  \href {https://doi.org/10.1088/1475-7516/2016/05/062}
  {\path{doi:10.1088/1475-7516/2016/05/062}}.

\bibitem{Khodadi:2017eim}
Mohsen Khodadi, Kourosh Nozari, and Anahita Hajizadeh.
\newblock {Some Astrophysical Aspects of a Schwarzschild Geometry Equipped with
  a Minimal Measurable Length}.
\newblock {\em Phys. Lett.}, B770:556--563, 2017.
\newblock \href {http://arxiv.org/abs/1702.06357} {\path{arXiv:1702.06357}},
  \href {https://doi.org/10.1016/j.physletb.2017.05.016}
  {\path{doi:10.1016/j.physletb.2017.05.016}}.

\bibitem{Scardigli:2018qce}
F.~Scardigli and R.~Casadio.
\newblock {Perihelion Precession and Generalized Uncertainty Principle}.
\newblock {\em Springer Proc. Phys.}, 208:149--155, 2018.
\newblock \href {https://doi.org/10.1007/978-3-319-94256-8_17}
  {\path{doi:10.1007/978-3-319-94256-8_17}}.

\bibitem{Tao:2012fp}
Jun Tao, Peng Wang, and Haitang Yang.
\newblock {Homogeneous Field and WKB Approximation In Deformed Quantum
  Mechanics with Minimal Length}.
\newblock {\em Adv. High Energy Phys.}, 2015:718359, 2015.
\newblock \href {http://arxiv.org/abs/1211.5650} {\path{arXiv:1211.5650}},
  \href {https://doi.org/10.1155/2015/718359} {\path{doi:10.1155/2015/718359}}.

\bibitem{Quintela:2015bua}
T.~S. Quintela, Jr., J.~C. Fabris, and J.~A. Nogueira.
\newblock {The Harmonic Oscillator in the Classical Limit of a Minimal-Length
  Scenario}.
\newblock {\em Braz. J. Phys.}, 46(6):777--783, 2016.
\newblock \href {http://arxiv.org/abs/1510.08129} {\path{arXiv:1510.08129}},
  \href {https://doi.org/10.1007/s13538-016-0457-9}
  {\path{doi:10.1007/s13538-016-0457-9}}.

\bibitem{Tkachuk:2013qa}
V.~M. Tkachuk.
\newblock {Deformed Heisenberg algebra with minimal length and equivalence
  principle}.
\newblock {\em Phys. Rev.}, A86:062112, 2012.
\newblock \href {http://arxiv.org/abs/1301.1891} {\path{arXiv:1301.1891}},
  \href {https://doi.org/10.1103/PhysRevA.86.062112}
  {\path{doi:10.1103/PhysRevA.86.062112}}.

\bibitem{Scardigli:2016pjs}
Fabio Scardigli, Gaetano Lambiase, and Elias Vagenas.
\newblock {GUP parameter from quantum corrections to the Newtonian potential}.
\newblock {\em Phys. Lett.}, B767:242--246, 2017.
\newblock \href {http://arxiv.org/abs/1611.01469} {\path{arXiv:1611.01469}},
  \href {https://doi.org/10.1016/j.physletb.2017.01.054}
  {\path{doi:10.1016/j.physletb.2017.01.054}}.

\bibitem{Zhao:2017xjj}
Qin Zhao, Mir Faizal, and Zaid Zaz.
\newblock {Short distance modification of the quantum virial theorem}.
\newblock {\em Phys. Lett.}, B770:564--568, 2017.
\newblock \href {http://arxiv.org/abs/1707.00636} {\path{arXiv:1707.00636}},
  \href {https://doi.org/10.1016/j.physletb.2017.01.029}
  {\path{doi:10.1016/j.physletb.2017.01.029}}.

\bibitem{Mu:2019bim}
Benrong Mu and Jun Tao.
\newblock {Minimal Length Effect on Thermodynamics and Weak Cosmic Censorship
  Conjecture in anti-de Sitter Black Holes via Charged Particle Absorption}.
\newblock 2019.
\newblock \href {http://arxiv.org/abs/1906.10544} {\path{arXiv:1906.10544}}.

\bibitem{Lu:2018mpr}
Fenghua Lu, Jun Tao, and Peng Wang.
\newblock {Minimal Length Effects on Chaotic Motion of Particles around Black
  Hole Horizon}.
\newblock {\em JCAP}, 1812:036, 2018.
\newblock \href {http://arxiv.org/abs/1811.02140} {\path{arXiv:1811.02140}},
  \href {https://doi.org/10.1088/1475-7516/2018/12/036}
  {\path{doi:10.1088/1475-7516/2018/12/036}}.

\bibitem{Hassanabadi:2019iff}
Hassan Hassanabadi, Elham Maghsoodi, and Won Sang~Chung.
\newblock {Analysis of motion of particles near black hole horizon under
  generalized uncertainty principle}.
\newblock {\em EPL}, 127(4):40002, 2019.
\newblock \href {https://doi.org/10.1209/0295-5075/127/40002}
  {\path{doi:10.1209/0295-5075/127/40002}}.

\bibitem{Maghsoodi:2020ura}
Elham Maghsoodi, Marc de~Montigny, Won~Sang Chung, and Hassan Hassanabadi.
\newblock {The effect of the Generalized Uncertainty Principle on the motion of
  particles near a black hole horizon}.
\newblock {\em Phys. Dark Univ.}, 28:100559, 2020.
\newblock \href {https://doi.org/10.1016/j.dark.2020.100559}
  {\path{doi:10.1016/j.dark.2020.100559}}.

\bibitem{Chen:2013pra}
Deyou Chen, Houwen Wu, and Haitang Yang.
\newblock {Fermion's tunnelling with effects of quantum gravity}.
\newblock {\em Adv. High Energy Phys.}, 2013:432412, 2013.
\newblock \href {http://arxiv.org/abs/1305.7104} {\path{arXiv:1305.7104}},
  \href {https://doi.org/10.1155/2013/432412} {\path{doi:10.1155/2013/432412}}.

\bibitem{Chen:2013tha}
Deyou Chen, Houwen Wu, and Haitang Yang.
\newblock {Observing remnants by fermions' tunneling}.
\newblock {\em JCAP}, 1403:036, 2014.
\newblock \href {http://arxiv.org/abs/1307.0172} {\path{arXiv:1307.0172}},
  \href {https://doi.org/10.1088/1475-7516/2014/03/036}
  {\path{doi:10.1088/1475-7516/2014/03/036}}.

\bibitem{Chen:2013ssa}
D.~Y. Chen, Q.~Q. Jiang, P.~Wang, and H.~Yang.
\newblock {Remnants, fermions` tunnelling and effects of quantum gravity}.
\newblock {\em JHEP}, 11:176, 2013.
\newblock \href {http://arxiv.org/abs/1312.3781} {\path{arXiv:1312.3781}},
  \href {https://doi.org/10.1007/JHEP11(2013)176}
  {\path{doi:10.1007/JHEP11(2013)176}}.

\bibitem{Chen:2014xgj}
Deyou Chen, Houwen Wu, Haitang Yang, and Shuzheng Yang.
\newblock {Effects of quantum gravity on black holes}.
\newblock {\em Int. J. Mod. Phys.}, A29(26):1430054, 2014.
\newblock \href {http://arxiv.org/abs/1410.5071} {\path{arXiv:1410.5071}},
  \href {https://doi.org/10.1142/S0217751X14300543}
  {\path{doi:10.1142/S0217751X14300543}}.

\bibitem{Maghsoodi:2019fca}
E.~Maghsoodi, H.~Hassanabadi, and Won Sang~Chung.
\newblock {Black hole thermodynamics under the generalized uncertainty
  principle and doubly special relativity}.
\newblock {\em PTEP}, 2019(8):083E03, 2019.
\newblock \href {http://arxiv.org/abs/1901.10305} {\path{arXiv:1901.10305}},
  \href {https://doi.org/10.1093/ptep/ptz085} {\path{doi:10.1093/ptep/ptz085}}.

\bibitem{Guo:2020xnf}
Xiaobo Guo, Kangkai Liang, Benrong Mu, Peng Wang, and Mingtao Yang.
\newblock {Chaotic Motion around a Black Hole under Minimal Length Effects}.
\newblock 2 2020.
\newblock \href {http://arxiv.org/abs/2002.05894} {\path{arXiv:2002.05894}}.

\bibitem{VERNER1996345}
J.H. Verner.
\newblock High-order explicit runge-kutta pairs with low stage order.
\newblock {\em Applied Numerical Mathematics}, 22(1):345 -- 357, 1996.
\newblock Special Issue Celebrating the Centenary of Runge-Kutta Methods.
\newblock URL:
  \url{http://www.sciencedirect.com/science/article/pii/S0168927496000414},
  \href {https://doi.org/https://doi.org/10.1016/S0168-9274(96)00041-4}
  {\path{doi:https://doi.org/10.1016/S0168-9274(96)00041-4}}.

\bibitem{lyapunov1992general}
Aleksandr~Mikhailovich Lyapunov.
\newblock The general problem of the stability of motion.
\newblock {\em International journal of control}, 55(3):531--534, 1992.

\bibitem{geist1990comparison}
Karlheinz Geist, Ulrich Parlitz, and Werner Lauterborn.
\newblock Comparison of different methods for computing lyapunov exponents.
\newblock {\em Progress of theoretical physics}, 83(5):875--893, 1990.

\bibitem{benettin1980lyapunov}
Giancarlo Benettin, Luigi Galgani, Antonio Giorgilli, and Jean-Marie Strelcyn.
\newblock Lyapunov characteristic exponents for smooth dynamical systems and
  for hamiltonian systems; a method for computing all of them. part 1: Theory.
\newblock {\em Meccanica}, 15(1):9--20, 1980.

\bibitem{Dalui:2018qqv}
Surojit Dalui, Bibhas~Ranjan Majhi, and Pankaj Mishra.
\newblock {Presence of horizon makes particle motion chaotic}.
\newblock {\em Phys. Lett.}, B788:486--493, 2019.
\newblock \href {http://arxiv.org/abs/1803.06527} {\path{arXiv:1803.06527}},
  \href {https://doi.org/10.1016/j.physletb.2018.11.050}
  {\path{doi:10.1016/j.physletb.2018.11.050}}.

\bibitem{Maldacena:2015waa}
Juan Maldacena, Stephen~H. Shenker, and Douglas Stanford.
\newblock {A bound on chaos}.
\newblock {\em JHEP}, 08:106, 2016.
\newblock \href {http://arxiv.org/abs/1503.01409} {\path{arXiv:1503.01409}},
  \href {https://doi.org/10.1007/JHEP08(2016)106}
  {\path{doi:10.1007/JHEP08(2016)106}}.

\bibitem{Zhao:2018wkl}
Qing-Qing Zhao, Yue-Zhou Li, and H.~Lu.
\newblock {Static Equilibria of Charged Particles Around Charged Black Holes:
  Chaos Bound and Its Violations}.
\newblock {\em Phys. Rev.}, D98(12):124001, 2018.
\newblock \href {http://arxiv.org/abs/1809.04616} {\path{arXiv:1809.04616}},
  \href {https://doi.org/10.1103/PhysRevD.98.124001}
  {\path{doi:10.1103/PhysRevD.98.124001}}.

\bibitem{poincare1881memoire}
Henri Poincar{\'e}.
\newblock M{\'e}moire sur les courbes d{\'e}finies par une {\'e}quation
  diff{\'e}rentielle (i).
\newblock {\em Journal de math{\'e}matiques pures et appliqu{\'e}es},
  7:375--422, 1881.

\bibitem{tabor1989chaos}
Michael Tabor.
\newblock {\em Chaos and integrability in nonlinear dynamics: an introduction}.
\newblock Wiley, 1989.

\bibitem{contopoulos2004order}
George Contopoulos.
\newblock {\em Order and chaos in dynamical astronomy}.
\newblock Springer Science \& Business Media, 2004.

\bibitem{Sekino:2008he}
Yasuhiro Sekino and Leonard Susskind.
\newblock {Fast Scramblers}.
\newblock {\em JHEP}, 10:065, 2008.
\newblock \href {http://arxiv.org/abs/0808.2096} {\path{arXiv:0808.2096}},
  \href {https://doi.org/10.1088/1126-6708/2008/10/065}
  {\path{doi:10.1088/1126-6708/2008/10/065}}.

\bibitem{Fischler:2015cma}
Willy Fischler and Sandipan Kundu.
\newblock {Hall Scrambling on Black Hole Horizons}.
\newblock {\em Phys. Rev. D}, 92(4):046008, 2015.
\newblock \href {http://arxiv.org/abs/1501.01316} {\path{arXiv:1501.01316}},
  \href {https://doi.org/10.1103/PhysRevD.92.046008}
  {\path{doi:10.1103/PhysRevD.92.046008}}.

\bibitem{Guo:2017bru}
Xiaobo Guo, Peng Wang, and Haitang Yang.
\newblock {Membrane Paradigm and Holographic DC Conductivity for Nonlinear
  Electrodynamics}.
\newblock {\em Phys. Rev. D}, 98(2):026021, 2018.
\newblock \href {http://arxiv.org/abs/1711.03298} {\path{arXiv:1711.03298}},
  \href {https://doi.org/10.1103/PhysRevD.98.026021}
  {\path{doi:10.1103/PhysRevD.98.026021}}.

\end{thebibliography}
\end{document}